# Translational diffusion in supercooled water at and near the glass transition temperature – 136 K

Greg A. Kimmel,* Megan K. Dunlap, Kirill Gurdumov, R. Scott Smith, Loni Kringle, and Bruce D. Kay

Physical Sciences Division, Pacific Northwest National Laboratory,

P.O. Box 999, Richland WA, USA 99352

*Corresponding author: gregory.kimmel@pnnl.gov

## ABSTRACT

The properties of amorphous solid water at and near the calorimetric glass transition temperature, $T_g$, of 136 K have been debated for years. One hypothesis is that water turns into a "true" liquid at $T_g$ (i.e., it becomes ergodic) and exhibits all the characteristics of an ergodic liquid, including translational diffusion. A competing hypothesis is that only rotational motion becomes active at $T_g$, while the "real" glass transition in water is at a considerably higher temperature. To address this dispute, we have investigated the diffusive mixing in nanoscale water films, with thicknesses up to ~100 nm, using infrared (IR) spectroscopy. The experiments used films that were composed of at least 90% $H_2O$ with $D_2O$ making up the balance and were conducted in conditions where H/D exchange was essentially eliminated. Because the IR spectra of multilayer $D_2O$ films (e.g., thicknesses of ~3 – 6 nm) embedded within thick $H_2O$ films are distinct from the spectrum of isolated $D_2O$ molecules within $H_2O$, the diffusive mixing of (initially) isotopically layered water films could be followed as a function of annealing time and temperature. The results show that water films with total thicknesses ranging from ~20 to 100 nm diffusively mixed prior to crystallization for temperatures between 120 and 144 K. The translational diffusion had an Arrhenius temperature dependence with an activation energy of 40.8 kJ/mol, which indicates that water at and near $T_g$ is a strong liquid. The measured diffusion coefficient at 136 K is $6.25 \pm 1.4 \times 10^{-21}$ m$^2$/s.

TOC Graphic

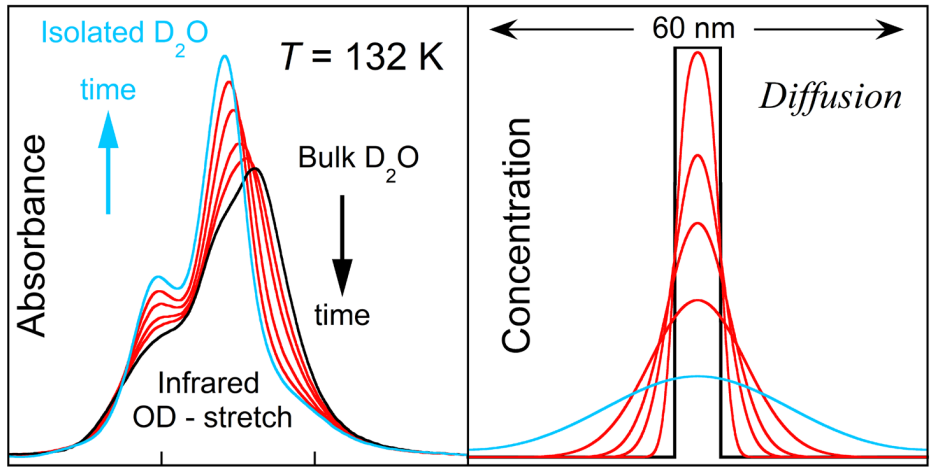

## I. INTRODUCTION

At sufficiently low temperatures (e.g., 100 K or less), water can adopt a variety of non-crystalline forms that are unable to relax on an experimentally accessible timescale. In an early report, Burton and Oliver found that depositing water vapor on a cold Cu rod at temperatures around 120 K produced an amorphous solid,[1] which is commonly called amorphous solid water (ASW). In 1985, Mayer showed that crystallization in liquid water could be avoided if micron-scale droplets were sprayed onto a cold substrate, producing hyperquenched glassy water (HGW).[2] Alternatively, Mishima and co-workers showed that compressing hexagonal ice, $I_h$, at 77 K produced amorphous ice,[3] that upon recovery to ambient pressure had a density of ~1.17 g/cm$^3$ and was thus called high density amorphous ice (HDA). Interestingly, upon annealing to ~135 K, HDA converts to a distinct polymorph – low density amorphous ice (LDA). When compressed, LDA reverts to HDA via a first-order-like transition.[4] At low pressures other amorphous solids, such as very high density and medium density amorphous ices (VHDA and MDA, respectively) have also been identified.[5, 6] The properties of amorphous solids at high pressures are also of great interest.

The possible connections between ASW, HGW, and LDA have been the extensively investigated.[7-9] One key question is: What happens to these amorphous solids as the temperature is increased to the point where the structure is no longer kinetically arrested?[9] One hypothesis is that they turn into the same supercooled liquid at 136 K, such that they exhibit all the properties characteristic of normal liquids, including rotational and translational diffusion, $D_{rot}$ and $D_{tr}$, respectively. (Below, we will refer to the translational diffusion of water molecules as simply "diffusion.") Using differential scanning calorimetry (DSC), a weak endotherm, which is found at ~136 K for LDA, ASW, and HGW, has been identified as the onset of liquid-like behavior. In that case, the corresponding glass transition temperature, $T_g$, would be 136 K.[10, 11] A second hypothesis is that this weak endotherm is associated with the unfreezing of rotational motion, but not diffusion, in these amorphous solids, and that the true glass-liquid transition occurs at much higher temperatures (e.g., ~165 K).[12, 13] Yet a third hypothesis is that the amorphous ices are unstable with respect to crystalline ice, and any observed changes are the result of an amorphous-to-crystalline transition. These and other hypotheses, are discussed in detail in several excellent reviews.[7-9]

One of the primary motivations for developing a detailed understanding of how amorphous ices transform when they are warmed up relates to persistent questions about the structure and dynamics of normal and supercooled liquid water, and its many anomalous properties.[7, 14, 15] A leading hypothesis proposes that liquid water at low temperatures and high pressures can exist in two distinct forms – a high-density liquid (HDL) and a low-density liquid (LDL) – that are separated by a two-phase coexistence line that terminates in a second critical point. One appealing aspect of this liquid-liquid critical point (LLCP) hypothesis is that it provides a natural explanation for HDA and LDA. Namely, they are the glassy analogs of HDL and LDL.[7] Therefore, determining if LDA turns into a "true" liquid at/near the glass transition temperature before it eventually crystallizes will provide valuable information about the feasibility of the LLCP and other hypotheses for water's unusual behavior.

Angell noted that "The most fundamental of the transport properties is the self-diffusion coefficient, since no external stress is required to manifest, or measure, the property."[16] The question of whether diffusion occurs in ASW, HGW, and LDA has been controversial since at least 1995 when Fisher and Devlin investigated H/D exchange in amorphous $H_2O$ films that had been doped with a low concentration of isolated $D_2O$.[12] Protons were injected into the films through photoexcitation of 2-napthol, and the H/D exchange kinetics, which first converted isolated $D_2O$ to 2 adjacent HOD ("coupled HOD") and then to isolated HOD, were monitored via infrared spectroscopy. Fisher and Devlin argued that the observed kinetics ruled out diffusion in ASW, and instead suggested molecular rotations were sufficient to explain their results. Subsequently, Johari challenged Fisher and Devlin's interpretation noting that diffusion could also account for the observations.[17] Recently, elegant experiments from Shephard and Salzmann examined the influence of isotopes ($H_2O$, $H_2^{18}O$ and $D_2O$) on the calorimetric glass transition temperatures in LDA, HDA and crystalline ice VI.[13] As they noted, hydrogen-disordered ice VI is an interesting case because its glass transition is due to unfreezing of molecular rotations – no diffusion is involved. Using DSC, they found that (i) the magnitude of endotherms for LDA and ice VI were comparable, (ii) their $T_g$'s were nearly the same, and (iii) the transition temperatures showed essentially identical isotopic shifts. Based on the similarities between LDA and ice VI, they concluded that the experimentally observed glass transition in LDA (and also HDA) involves only rotational motion.

Vapor deposition onto cold surfaces in ultrahigh vacuum with various isotopologues of water allows one to create isotopically-layered amorphous solid water films.[18] Upon heating, such layered films have been used to investigate dynamic processes, including mixing within nanoscale water films.[19, 20] The mixing within layered $H_2^{18}O/H_2^{16}O$ films, which was monitored via desorption into the gas phase, occurred in concert with crystallization of the films at ~ 155 K. A model that treated the crystallization kinetics and assumed diffusion within in the liquid portion of the crystallizing films (with negligible diffusion in the crystalline portion) was able to reproduce the observations. However, subsequent experiments showed that the crystallization heavily influenced the mixing process, precluding an accurate determination of the diffusion in the liquid portion from those measurments.[21, 22]

Fluidity – which is connected to diffusion – is one of the hallmarks of liquids. In a recent review of "Water's controversial glass transitions", Amann-Winkel, et al., note that "(t)he key question, to us, remains whether above the glass transition the water molecules display liquid-like bulk fluidity or not."[9] However, many of the experimental approaches used to date do not directly address this issue or had various experimental limitations. The authors suggest that detailed measurements of the shear viscosity is one approach to this problem, while "diffusion measurements probing the transport of *oxygen* [emphasis in the original] will do the job."

Here, building upon recent experiments showing that H/D exchange can be effectively eliminated in nanoscale water films,[23] we use infrared spectroscopy to investigate the diffusive mixing of (intact) $D_2O$ molecules in $H_2O$ films at and below the traditional glass transition ($T_g$ = 136 K). The results demonstrate long range translational diffusion (e.g. > 5 nm) of molecular $D_2O$ in water at temperatures from 120 - 144 K. The diffusive mixing of the films is independent of the total film thicknesses, $x_{film}$, in the range of ~20 – 100 nm. The results indicate that the measured diffusion is not influenced by the water/substrate or water/vacuum interfaces and is instead characteristic of the bulk liquid. The diffusion is activated with simple Arrhenius temperature dependence and an activation energy of 40.8 kJ/mol. The results indicate that water at and near the glass transition is a strong, supercooled liquid. Furthermore, the Wilson-Frenkel model[24-26] – which posits that the growth rate of a crystalline phase in contact with its melt is proportional to the diffusion coefficient within the liquid – holds for supercooled water near $T_g$.

## II. EXPERIMENTAL METHODS

The experiments were performed in an ultrahigh vacuum (UHV) system, which had typical base pressures of $1.3 \times 10^{-8}$ Pa or less, that has been described in detail previously.[27] For the results reported here, the relevant components were a closed-cycle helium cryostat (Advanced Research Systems, CSW-204B), an effusive molecular beam dosing line, a quadrupole mass spectrometer (Extrel, Merlin) and a Fourier transform infrared spectrometer (Bruker, Vertex 70). The cryostat allowed a Pt(111) single crystal (1 cm diameter, 2 mm thick) to be cooled to a base temperature of ~25 K. Heating and isothermal temperature control were achieved by resistively heating thin tantalum wires spot-welded to the back of the crystal. The temperature was monitored with a K-type thermocouple, also spot-welded to the back of the crystal.

The Pt(111) crystal was cleaned by sputtering with 2 keV Ne$^+$ and then annealing at 1000 K in vacuum. Nanoscale films of H$_2$O and D$_2$O were adsorbed onto the crystal at normal incidence at 108 K using the molecular beam with fluxes of ~2 × 10$^{18}$ m$^{-2}$s$^{-1}$. These conditions produced a non-porous, amorphous solid water (ASW) film.[28, 29] The central portion of the molecular beam (the umbra) had a diameter of 8.5 mm, while at the edge of the crystal, the flux decreased by ~20% (i.e., in the penumbra). Water coverages, θ, are given in units of water monolayers (ML) on Pt(111), which was determined using temperature programmed desorption (TPD). For a water "monolayer" on Pt(111), there are two closely related structures corresponding to a single monolayer on Pt(111), the $(\sqrt{37} \times \sqrt{37})$R25.3° structure and the $(\sqrt{39} \times \sqrt{39})$R16.3° structure.[30] They have coverages of 1.054 and 1.077 × 10$^{15}$ molecules/cm$^2$, respectively and are difficult to distinguish in the TPD spectra, but this uncertainty is small compared to other sources of error in the measurements. Below, some results are discussed in terms of the thickness of various water layers. To convert water coverages to thicknesses, we assumed 1 ML = 0.33 nm. (For LDA at 80 K, the density is 937 kg/m$^3$,[31] and the typical distance between molecules $r_{nn}$ can be estimated as $r_{nn} \sim N^{-1/3} \sim$ 0.32 nm, where $N$ is the number density.)

Water films with a total coverage, θ$_{total}$, of 200 ML, corresponding to a thickness of ~66 nm, were used for most of the experiments presented below. However, as shown in the Results and Discussion section, the same behavior was found for films with coverages from 60 – 300 ML, corresponding to thicknesses of ~20 to 100 nm. Furthermore, in most of the experiments, the D$_2$O layers were sufficiently far from the water/Pt and water/vacuum interfaces that processes occurring at those interfaces did not influence the results. The disruption of the bulk hydrogen-bonding network in the vicinity of these interfaces could alter the structure and dynamics there, making them unrepresentative of the bulk transport properties. For example, crystallization in nanoscale water films occurs preferentially at the vacuum interface, presumably due to the enhanced mobility and excess free volume for water molecules there.[32] In another example, the enhanced mobility of molecules at the vacuum interface can lead to the formation of exceptionally stable glasses during vapor deposition.[33, 34]

To investigate molecular diffusion using layered films of D$_2$O and H$_2$O it was necessary to suppress H/D exchange. As mentioned already in the introduction, H/D exchange can convert D$_2$O into HOD. In that case, repeated H/D exchange reactions combined with molecular rotations, can lead to transport of hydrogenic mass over appreciable distances, even in the absence of molecular diffusion. Because the relative rates for molecular rotations and diffusion are not known near $T_g$, experiments with appreciable H/D exchange do not provide an unambiguous method for measuring diffusion.[9] However, at temperatures near $T_g$, autoionization in water is very low, and recent work in our group has shown that H/D exchange can be suppressed if exogenous sources of excess protons are removed.[23] The primary source of protons for the experiments reported here was from dissociative adsorption of $H_2$ on the Pt(111) substrate that occurred as the sample cooled after rapid heating to high temperatures. (The "flash" heating was performed to remove any volatile species from the surface prior to adsorbing the water films. $H_2$ is typically one of the primary residual gases in ultrahigh vacuum systems.) Once water was deposited and the system was heated, the adsorbed H atoms reacted with the water to form hydrated protons that subsequently diffuse into the film, leading to H/D exchange between $D_2O$ and $H_2O$. However, this problem was avoided by adsorbing small amounts of $O_2$ on the surface prior to adding the water. The oxygen effectively scavenged the adsorbed H, suppressing the H/D exchange but not eliminating it entirely.[23] However, any remaining excess protons established a distance-dependent distribution within the film, such that they were primarily localized near the Pt substrate.[27] Because the IR signal for isolated $D_2O$ is easily distinguished from the signal for an isolated HOD,[12, 27] it was relatively straightforward to monitor the production of isolated HOD (if any) during the course of the experiments. For example, in some experiments where the evolution of the water films was monitored for long times, $D_2O$ diffused into the vicinity of the Pt substrate and some H/D exchange occurred.

    Water films in UHV systems are metastable with respect to both crystallization and desorption (sublimation/vaporization). The experiments reported below were designed to probe processes in water films at/near $T_g$ that were not affected by crystallization or desorption. In IRAS, crystallization can typically be detected once the crystalline fraction has reached ~ 0.01 – 0.02.[35, 36] The results presented below focus on times prior to the onset of crystallization. For T < 136.5 K, no crystallization was detected during the experiments. However, for some experiments at higher temperatures, the films eventually crystallized at longer times. Only data without detectable crystallization was included in the analysis. Furthermore, the times most relevant for analyzing the diffusion were considerably less than the crystallization time. For example, the onset of crystallization was observed at ~2700 s for a film annealed at 140 K. Below we will show that a typical water molecule would have diffused $x_{diff}$ ~ 16 nm during that time (where $x_{diff} \sim \sqrt{6D_{tr}t}$ (in 3 dimensions)). The experiments started with $D_2O$ films, $\theta_{D2O}$ = 2 – 20 ML, deposited at various locations within $H_2O$ films (e.g., **Fig. 1a**). Desorption from the water films did not appreciably affect the results for experiments where the coverage of an $H_2O$ cap layer, $\theta_{cap}$, deposited on top of the $D_2O$ layer was large compared to amount that desorbed during the experiment.

Those experiments included films with $\theta_{total}$ = 200 ML, a 20 ML D$_2$O layer, and $\theta_{cap}$ = 90 ML. For experiments with D$_2$O layers at or near the vacuum interface, desorption did not qualitatively change the results. However, to obtain quantitative results, desorption was included in the analysis.

In the results presented below, the OD-stretch region (~2200 cm$^{-1}$ to 2750 cm$^{-1}$) of the IRAS spectra was analyzed to assess the diffusive mixing in D$_2$O/H$_2$O water films. In this wavenumber range, H$_2$O has a broad combination band (see **Fig. S1**). The contribution of this H$_2$O band has been subtracted prior to analysis and displaying the spectra.

To model diffusion in the water films, the one-dimensional diffusion equation was solved by converting it into a series of coupled ordinary differential equations (representing the layers within the films). The initial conditions were chosen to match the experimental configurations in the layered films of H$_2$O and D$_2$O (e.g., **Fig. 1a**). A reflecting boundary condition was imposed at the water/Pt and water/vacuum interfaces. For some experiments, D$_2$O layers were deposited at or near the vacuum interface. As noted above, desorption was not negligible for those experiments, and it was included in the calculations. Although the desorption rate for films of pure D$_2$O and H$_2$O are different at a given temperature, the D$_2$O and H$_2$O were assumed to desorb at the same rate in the simulations. Test calculations, which varied the desorption rate, indicated that this did not appreciably influence the results.

## III. RESULTS

To address molecular translational diffusion in "bulk" water, the experiments used infrared reflection absorption spectroscopy (IRAS) to monitor the evolution versus time in the OD-stretching region of water films that were deposited with layers of D$_2$O embedded within H$_2$O in various configurations (e.g., **Fig. 1a**). The water layers were grown in conditions where diffusion was negligible and the initial concentration of D$_2$O in the film was zero except in well-defined layers. If molecular diffusion is appreciable, then upon annealing to higher temperature, the overall concentration of D$_2$O will evolve towards a constant determined by the relative amounts of H$_2$O and D$_2$O within the film. IRAS can monitor this process because IR spectra in the OH- and OD-stretching regions are sensitive to the local hydrogen bonding arrangement.[37] Here, we use the fact that D$_2$O molecules isolated within an H$_2$O matrix have an IR spectrum in the OD-stretch region (~2200 – 2750 cm$^{-1}$) that is distinct from those of both pure D$_2$O and isolated HOD (see **Fig. S2**).[12, 27]

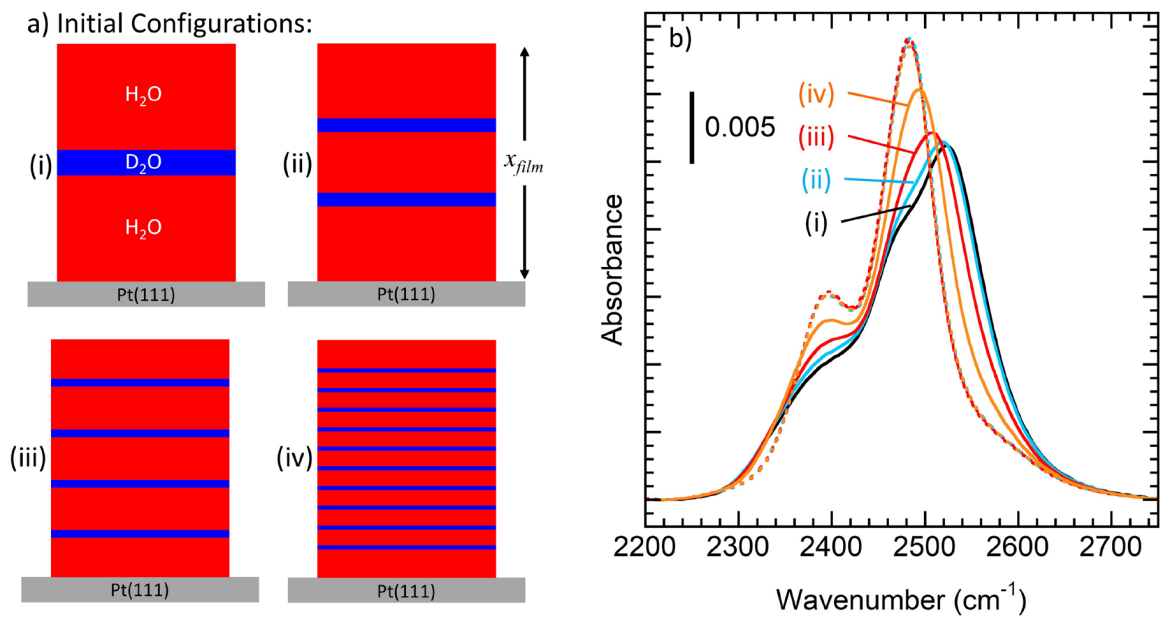

**Fig. 1.** (a) Schematic of the initial sample configurations. All the films were comprised of 180 ML $H_2O$ and 20 ML $D_2O$ arranged in four different configurations: (i) A single 20 ML $D_2O$ layer embedded in the middle of the $H_2O$; (ii) 2 × 10 ML $D_2O$ layers; (iii) 4 × 5 ML $D_2O$ layers, and (iv) 10 × 2 ML $D_2O$ layers. (b) IRAS spectra for 4 water films with different initial configurations: As deposited (solid lines) and during (dotted lines) annealing at 134.5 K for (i) $2.1 \times 10^4$ s, (ii) 5270 s, (iii) 1330 s, and (iv) 209 s. The IR spectra before (during) annealing were taken at 108 K (134.5 K). After annealing all the spectra are essentially identical (dotted lines).

**Figure 1b** shows several spectra for water films that had $\theta_{total}$ = 200 ML (i.e., $x_{film}$ ~ 66 nm), with $\theta_{H2O}$ = 180 ML and $\theta_{D2O}$ = 20 ML. While the films all contained 20 ML $D_2O$, the initial spatial arrangement of the $D_2O$ was different in each film (**Fig. 1a**). As a result, the IR spectra before annealing were all distinct (**Fig. 1b**, solid lines). The IR spectra of films grown with fewer, but thicker, $D_2O$ layers more closely resemble the spectra of "bulk" $D_2O$ because they have fewer $D_2O$ molecules influenced by nearby $H_2O$ (see **Fig. S2**). In contrast, after annealing for various times at 134.5 K, all the films evolved to the point where they had a nearly identical spectrum that was dominated by two distinct peaks at 2478 cm$^{-1}$ ± 2 cm$^{-1}$ and 2392 cm$^{-1}$ ± 2 cm$^{-1}$ (**Fig. 1b**, dotted lines). IR spectra were also obtained for films with lower $D_2O$ concentrations dispersed in $H_2O$ (see **Fig. S3**), and they were similar to the final spectra shown in **Fig. 1**. The IR spectra observed after sufficient annealing are characteristic of isolated $D_2O$ and are quite similar to previous reports.[12, 23, 27] Note that as the thickness of the individual $D_2O$ layers increased in **Fig. 1**, the annealing time required for the resulting IR spectra to resemble isolated $D_2O$ increased rapidly. For film (i), which had a single 20 ML $D_2O$ layer, to sufficiently mix to have a similar spectrum to the other starting configurations suggests that water molecules within that film diffused distances that were on the order of 10 nm.

For the experiments shown in **Fig. 1**, 10% of the water molecules were $D_2O$ while the remaining 90% were $H_2O$. At this concentration, if all the water molecules form 4 hydrogen bonds and are randomly distributed with respect to the isotopologues, then 66% of the $D_2O$ will have 4 $H_2O$ neighbors and another 29% will have only 1 $D_2O$ neighbor. As a result, isolated $D_2O$ and $D_2O$ "dimers" will account for 95% of the total $D_2O$ in the water film. Here, we refer to low concentrations of $D_2O$ dispersed in $H_2O$ as "isolated $D_2O$", but it is useful to remember that the actual amount of $D_2O$ nearest neighbors (or other $D_2O$ clusters) is a sensitive function of the local concentration.

For experiments with a single multilayer $D_2O$ slab in the middle of an $H_2O$ film (see **Fig. 1a(i)**), the IR spectra showed a characteristic evolution versus time when the films were annealed at temperatures where diffusion was appreciable. For example, **Fig 2a** shows a series of IR spectra in the OD-stretch region for a film with one 20 ML $D_2O$ layer in the middle of 180 ML $H_2O$. Before annealing, the peak in the spectrum was at ~2490 cm$^{-1}$ (**Fig. 2a**, black line). Upon annealing at 132.5 K, the peak continuously shifted to lower wavenumbers (**Fig 2a**, red lines). At the same time, an initial shoulder at ~2395 cm$^{-1}$ developed into a distinct peak at later times. The changes in the IR spectra upon annealing can be highlighted by taking the difference between the spectra at any given time and the first spectrum at $T_{anneal}$ (**Fig. 2b**). The difference spectra emphasize the emergence of the 2 lower frequency peaks in the IR spectra at later times.

For water, the IR spectra in the OH- and OD-stretch regions are very sensitive to the local environment,[37] including the local concentration of $H_2O$ and $D_2O$. They also reflect the sum of the contributions from all the molecules. For experiments such as those shown in **Figs. 1** and **2**, the $D_2O$ molecules experience a time-dependent range of local concentrations, which makes it difficult to extract the self-diffusion coefficient for water directly from the observed IR spectra. However, for films with 20 ML $D_2O$ in the middle of 180 ML $H_2O$ that were subsequently annealed at different temperatures, the IR spectra evolved through essentially the same sequence as shown in **Fig. 2**, but the amount of time needed to progress through the sequence depended on the temperature. Qualitatively, this is just what one expects for a system that diffusively mixes from an initially layered configuration. In that case, there is a characteristic time at each temperature, $\tau(T)$, that is proportional to $\lambda^2/D_{tr}(T)$, where $\lambda$ is characteristic length within each film. For films with the same initial configuration, $\lambda$ does not depend on temperature, so the time dependence observed in the experiments is related to $D_{tr}(T)$. While it is difficult to determine $D_{tr}(T)$ directly from the IR spectra, we demonstrate next that it is straightforward to determine $\tau(T)$ (e.g., **Fig. 3**). Another set of experiments (discussed below), then allow us to relate $\tau(T)$ to $D_{tr}(T)$.

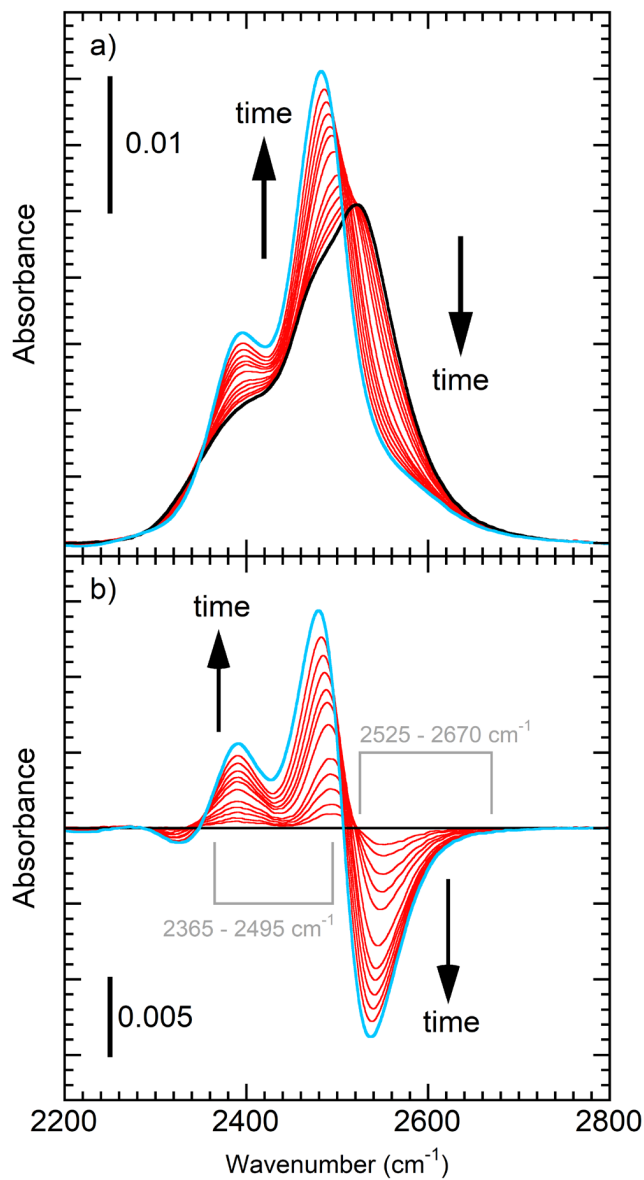

**Fig. 2**. a) IR spectra in the OD-stretch region for a film that started with 20 ML $D_2O$ deposited in the middle of 180 ML $H_2O$ (see, Fig. 1a(i)). All the spectra were acquired at 132.5 K during annealing. The first spectrum at 132.5 K (black line) is similar to the as-deposited spectrum acquired at 108 K (not shown). The spectrum after annealing for $4.4 \times 10^4$ s (blue line) is characteristic of isolated $D_2O$ in $H_2O$. The intermediate spectra (red lines) were taken at 82, 164, 370, 660, 1080, 2650, 4510, 6020, 9040, and $1.38 \times 10^4$ s. b) Difference spectra where the first spectrum (panel a, black line) has been subtracted from all the subsequent spectra. With increasing annealing time (red lines), the spectra lose intensity at higher wavenumbers (e.g., 2525 – 2670 $cm^{-1}$) and gain intensity at lower wavenumbers (e.g. 2365 – 2495 $cm^{-1}$). Fig. S1 shows the raw spectra.

To assess the timescales for the changes in the IR spectra due to diffusive mixing of the water layers at various temperatures, we can track integrals over various portions of the IR bands versus time. **Figs. 3a** and **S4a** show two of these integrals: one on the high-frequency side of the OD-stretch band (2525 – 2670 cm$^{-1}$, see **Fig. 2b**) that decreases with time, and a second integral on the low-frequency side (2365 – 2495 cm$^{-1}$, see **Fig. 2b**) that increases with time. As seen in **Figs. 3a** and **S4a**, the time at which a given value of either integral is reached increases substantially at lower temperatures. However, when the times are scaled by $\tau(T)$, the data collapse onto two curves – one for each of the integrals (**Figs. 3b** and **S4b**). An important observation is that $\tau(T)$ increased exponentially versus $1/T$ with an apparent activation energy of 40.8 kJ/mol (see **Fig. S5** and **Table S1**). Section S1 of the **supplementary material** discusses the scaling behavior expected for diffusively mixed films and our method for determining $\tau(T)$.

Qualitatively similar results to those shown in **Fig. 3** were obtained for different choices of the integration limits within the increasing or decreasing portions of the IR band. As noted already, this behavior is consistent with diffusive mixing in the films where only the rate of mixing – not the sequence of concentration profiles – depends on the temperature. Similar scaling behavior was also observed for other film geometries (see **Fig. S6**). For the results shown in **Fig. 3**, the IR spectra continued to evolve even at the longest times in part because the D$_2$O was still not uniformly distributed within the water films. While the D$_2$O should continue to disperse at longer times, several factors worked against conducting longer experiments, including desorption and crystallization of the water films, and an increase in the H/D exchange when D$_2$O diffused into the vicinity of the platinum substrate (see discussion in the Methods section).

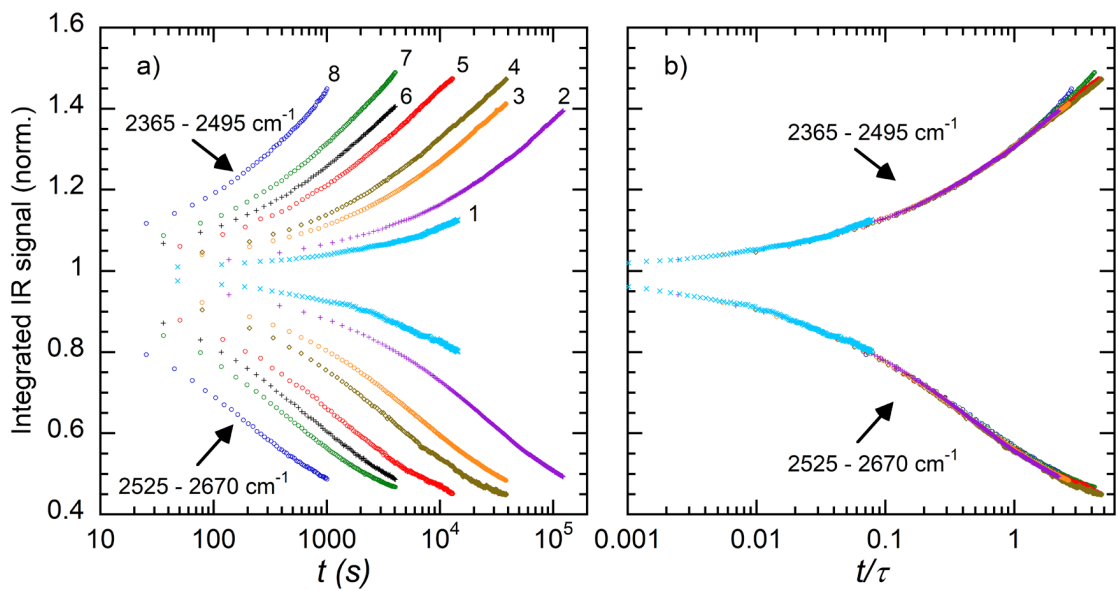

**Fig. 3.** (a) Integrals over low wavenumber (2365 – 2495 cm$^{-1}$) and high wavenumber (2525 – 2670 cm$^{-1}$) regions of the OD-stretch band vs time. Water films with one 20 ML D$_2$O layer in the middle of 180 ML H$_2$O were annealed at (1) 120.2 K, (2) 124.5 K, (3) 128.5 K, (4) 130 K, (5) 134 K, (6) 136 K, (7) 138.5 K, and (8) 142.5 K. Each symbol in the figure corresponds to an IR spectrum taken at the indicated time and temperature. The increasing signal in the low wavenumber portion of the band is associated with increasingly isolated D$_2$O, while the decreasing signal at higher wavenumbers is due to the loss of "bulk-like" D$_2$O. b) When the times in a) are scaled by $\tau(T)$, all the data collapse onto 2 curves, as expected for diffusion. (**Fig. S7** displays this data on a linear time scale.)

One concern that has been raised with respect to experiments on nanoscale water films is that they might not represent the behavior of bulk liquid water. While numerous classical and *ab initio* molecular dynamics simulations have suggested that the structure and dynamics of water converge to bulk behavior in very short distances (typically less than 3 nm) from water/solid, water/vacuum, or water/air interfaces,[38-40] it is important to investigate if this also holds for water films near $T_g$. To test this, diffusive mixing was measured in a series water films with a 10 ML D$_2$O layer deposited between H$_2$O cap and spacer layers of increasing thickness separating the D$_2$O from the interfaces (i.e. similar to **Fig. 1a(i)**). Specifically, the as-deposited film structures were Pt/$\theta_{spacer}$/10 ML D$_2$O/$\theta_{cap}$, where 10 ML $\leq \theta_{spacer} = \theta_{cap} \leq$ 150 ML. The corresponding thicknesses for these films were ~10 nm $\leq x_{film} \leq$ 100 nm. For these experiments, if the diffusion was independent of the film thickness, then the concentration vs time would be the same for all the different geometries until D$_2$O diffused to the water/Pt and water/vacuum interfaces. For water films annealed at 134 K for 8000 s, the IR spectra were essentially identical for H$_2$O layers $\geq$ 45 ML (see **Fig. S8**). For thinner H$_2$O cap and spacer layers, the IR spectra begin show differences at early times because D$_2$O reached the interfaces sooner. Based on these results, the diffusion coefficient in the nanoscale water films was independent of thickness for at least $x_{film} > 20$ nm.

While it was difficult to determine $D_{tr}(T)$ when the D$_2$O layers were embedded in the middle of thick H$_2$O films, it could be measured with experiments for which D$_2$O layers were deposited at or near the vacuum interface. **Fig. 4a** shows IR spectra for an experiment where a 20 ML D$_2$O layer was adsorbed on top of a 180 ML H$_2$O film and then annealed at 132.5 K. Initially (black line), the spectrum had a weak, narrow peak at 2727 cm$^{-1}$ due to non-hydrogen bonded OD groups ("dangling ODs") of D$_2$O molecules at the vacuum interface.[41, 42] At later times, D$_2$O diffused into the H$_2$O layer such that the total concentration of D$_2$O at the vacuum interface decreased. As a result, the dangling OD signal gradually decreased (**Fig. 4a**, red and blue lines). As the dangling OD signal decreased, the main OD-stretch band also evolved

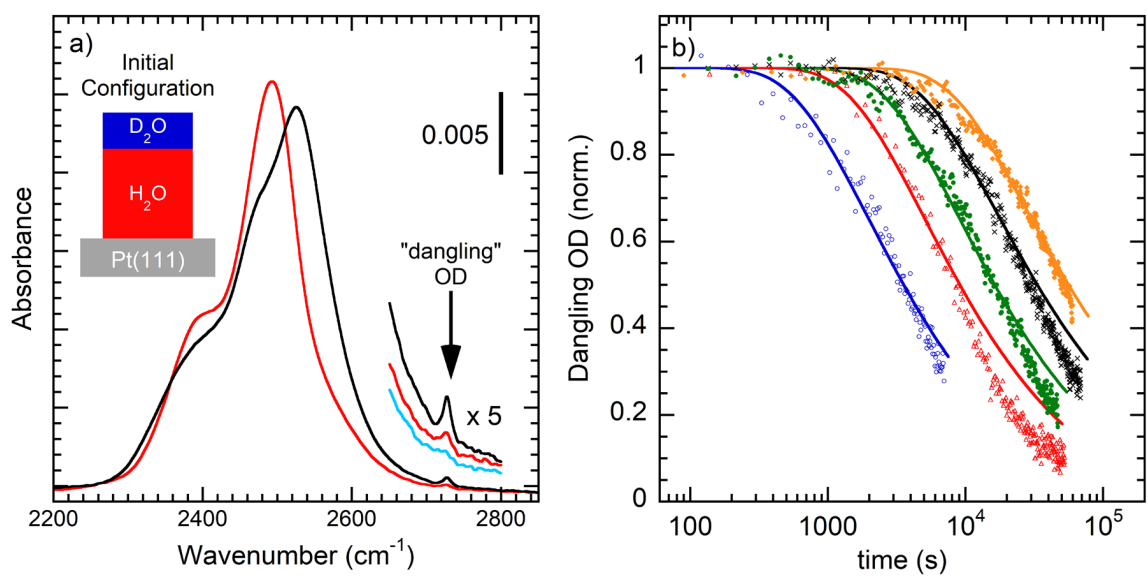

**Fig. 4.** a) IR spectra for a 20 ML D$_2$O film deposited on top of 180 ML H$_2$O and annealed at 132.5 K. As deposited, the spectrum has a small peak at 2727 cm$^{-1}$ that is due to non-hydrogen-bonded OD groups at the water/vacuum interface (black line). After annealing for $1.9 \times 10^4$ s, the OD-stretch band evolves towards a shape characteristic of isolated D$_2$O and the intensity of the dangling OD-peak decreases (red line). At even longer times, the dangling OD signal continues to decrease (e.g., $3.9 \times 10^4$ s, blue line). b) Integrated intensity of the dangling OD peak (symbols) versus time for films annealed at 128.5 (orange diamonds), 130.5 (black crosses), 132.5 (green circles), 134 (red triangles), and 138 K (blue circles). The solid lines show the surface concentration calculated for diffusion coefficients of $7.7 \times 10^{-22}$ (orange), $1.3 \times 10^{-21}$ (black), $2.3 \times 10^{-21}$ (green), $4.3 \times 10^{-21}$ (red) and $1.1 \times 10^{-20}$ m$^2$/s (blue).

---

towards the spectrum characteristic of isolated D$_2$O in H$_2$O. We assume that the dangling OD signal was proportional to the fraction of D$_2$O in the layer at the vacuum interface. In that case, the signal, normalized by its value when the water at the interface was entirely D$_2$O, gave the relative concentration of D$_2$O at the interface. **Fig. 4b** (symbols) shows the normalized integrated intensity for the dangling OD signal versus time for films annealed at 128.5 – 138 K. The solid lines show corresponding calculations of the concentration of D$_2$O at the vacuum interface for diffusion coefficients that range from $7.7 \times 10^{-22}$ m$^2$/s to $1.1 \times 10^{-20}$ m$^2$/s. Because desorption from the films was appreciable on the timescale of the experiments, it was also included in the simulation. For example, ~ 6 ML D$_2$O and 9 ML H$_2$O desorbed during the experiment at 138 K (**Fig. 4b**, blue circles). **Fig. S9a** shows the data in Fig. 4b, along with several more experiments at other temperatures. All the results, collapse onto a common curve when the times are scaled by $\tau(T)$ (**Fig. S9b**).

A separate set of experiments, which tracked the evolution of the dangling OD signal vs time at 138 K for a D$_2$O layer deposited at or near the vacuum interface, was also used to quantify the diffusion (see **Fig. 5**). In these experiments, the dangling OD signal was initially zero when the D$_2$O layer was capped with H$_2$O (e.g., **Fig. 5a.**, red triangles). As the D$_2$O diffusively mixed with the H$_2$O, the D$_2$O at the interface increased at early times before decreasing again at longer times. An example of the calculated concentration profiles for a D$_2$O layer with θ$_{cap}$ = 5 ML at several times is shown in **Fig. 5b**. Overall, the calculations reproduce the trends observed in data (**Fig. 5a**, dotted lines). However, the concentration at the interface deduced from the dangling OD signal was consistently 5 – 10% lower than the calculations. In the figure, the calculated concentrations have been scaled by 0.90 for θ$_{cap}$ = 3 – 20 ML and 0.95 for θ$_{cap}$ = 1 ML. (**Fig. S10** shows the results without the adjustment, and **Fig. S11** shows how changing $D_{tr}(T) \pm$ 30% affects the calculations.) This discrepancy could be related to the fact the dangling OD signal was very weak compared to the OD-stretch band, which resulted in a noisier signal and larger uncertainty in its normalization. However, the IR spectra also showed that there was a small amount of H/D exchange at the vacuum interface that led to the production of some isolated HOD. This effect, which was negligible for D$_2$O layers located in the middle of thick H$_2$O films, was not included in the calculation.

The red squares in **Fig. 6** show the diffusion coefficients determined from measurements of the dangling OD versus $1/T$. As seen in the figure, the diffusion coefficient decreases rapidly as the temperature decreases. For the experiments shown in **Fig. 3**, the characteristic times, $\tau(T)$, increased exponentially as the temperature decreased (**Fig. S5**). As mentioned already, it is difficult to extract $D_{tr}(T)$ from those measurements. However, the experiments with D$_2$O layers in the middle of films with different overall thicknesses (e.g., **Fig. S8**) and with the D$_2$O layers deposited at different heights within films of the same thickness (e.g., **Fig. 5**) indicate that the diffusion is largely independent of the location within these water films for a wide range of film thicknesses. In that case, a single constant, $\lambda^2$, relates $\tau(T)$ to the diffusion coefficient: $D_{tr}(T) = \lambda^2/\tau(T)$. Assuming $\lambda$ = 3.3 nm (see **supplementary materials** section S1), the black diamonds in **Fig. 6** show $D_{tr}(T)$ calculated from the characteristic times. The results show that the diffusion coefficient increased by a factor of ~900 when the temperature increased from 120.2 to 144.5 K. The dashed line shows an Arrhenius fit to the data with an activation energy, $E_a$, of 40.76 kJ/mol and a prefactor of 2.84 × 10$^{-5}$ m$^2$/s.

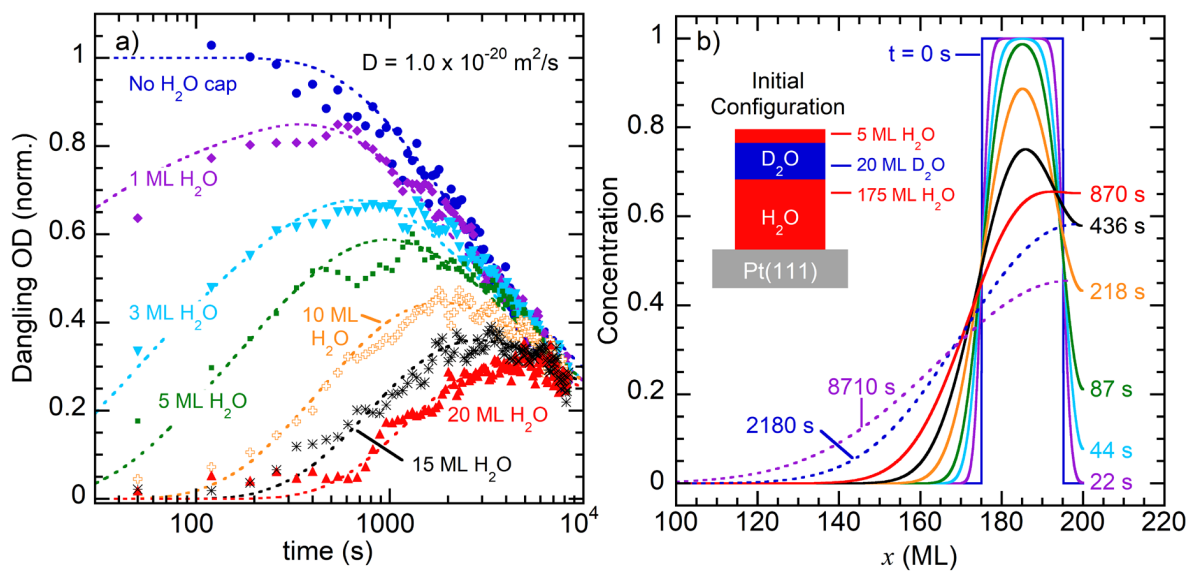

**Fig. 5.** a) Normalized integrated intensity of the dangling OD peak (symbols) versus time for water films with 20 ML D$_2$O and 180 ML H$_2$O annealed at 138 K. The D$_2$O layers were capped with H$_2$O with coverages, θ$_{cap}$, of 0, 1, 3, 5, 10, 15, and 20 ML. The dotted lines show the calculated concentration of D$_2$O at the vacuum interface assuming $D_{tr}$ = 1.0 × 10$^{-20}$ m$^2$/s. b) Calculated concentration profiles vs position ($x$) within the film at several times for a water film with θ$_{cap}$ = 5 ML. The calculations, which include desorption from the film, illustrate the initial increase and eventual decrease in the concentration of D$_2$O at the vacuum interface.

For several liquids, experiments have shown that the growth rate of a crystal in contact with its melt is proportional to the diffusion rate in the (supercooled) liquid.[43, 44] The growth rate, $G(T)$, can be decomposed into kinetic and thermodynamic components.[24-26] When the thermodynamic driving force for crystallization is large, the Wilson-Frenkel model can be used to describe the growth rate: $D_{tr}(T) = \alpha G(T)/[1 - \exp(-\Delta G_{lx}(T)/k_b T)]$, where $\Delta G_{lx}(T)$ is the free energy difference between the crystal and liquid and $\alpha$ is a constant related to the width of the liquid/solid interface and the length of a diffusive step in the liquid. Previous work in our group measured the growth rate of crystalline ice for temperatures between 125 and 260 K.[45] It assumed that the Wilson-Frenkel model held for supercooled water and used it to extract the diffusion in the liquid. In those experiments, α was determined by comparing the measured $G(T)$ near 250 K with independent measurements of $D_{tr}(T)$ in $H_2O$. That value of α was then used to predict the $D_{tr}(T)$ for 125 K ≤ T ≤ 260 K. With the independent measurements of the diffusion rates presented here and those earlier growth rates, we can now test the validity of the Wilson-Frenkel model at temperatures near $T_g$: The blue circles in **Fig. 6** show the diffusion rates previously estimated from the ice growth rates for T ≤ 152 K along with the current diffusion data. As seen in the figure, both the overall magnitude and the activation energy for $D_{tr}(T)$ predicted by the Wilson-Frenkel model are comparable to those measured here.

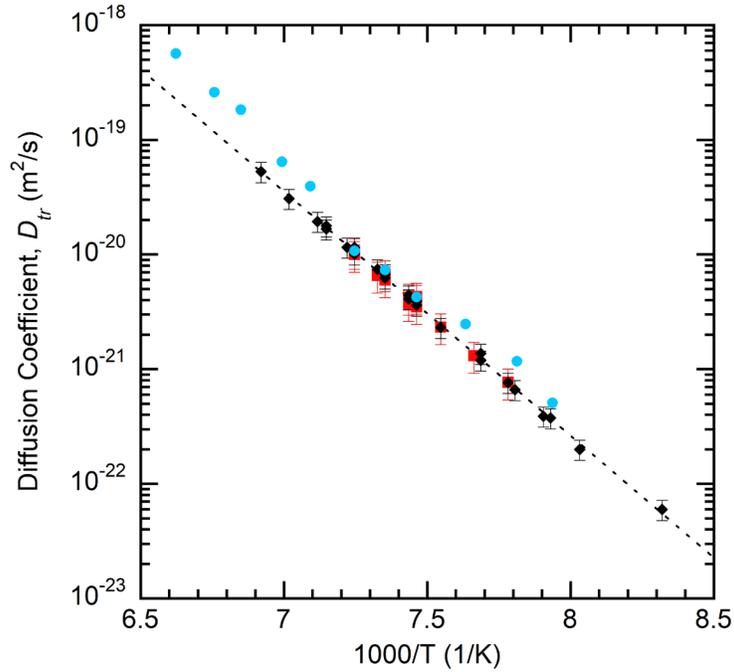

**Fig. 6.** Self-diffusion coefficient for water, $D_{tr}(T)$, vs $1000/T$. The red squares show the diffusion coefficient determined from measurements of the dangling OD signal versus annealing time (see **Figs. 4 and 5**). The black diamonds show $D_{tr}(T)$ determined from the characteristic diffusion times, τ, for films with 20 ML $D_2O$ embedded within 180 $H_2O$ (see **Fig. S5**). The blue circles show previous estimates of $D_{tr}(T)$ using measurements of the crystalline ice growth rate and the Wilson-Frenkel model.[45]

## IV. DISCUSSION

In many supercooled liquids, the structural relaxation times and viscosity increase very rapidly near $T_g$, without substantial changes in the structure.[46, 47] Liquids with this property are called "fragile", and it is generally believed that the fragility is related to the development of dynamic heterogeneity within the supercooled liquids.[46, 47] In contrast to fragile liquids, "strong" liquids exhibit Arrhenius temperature dependence in their dynamic properties near $T_g$. Above the water's melting point, the diffusion and viscosity follow approximately Arrhenius temperature dependences. At lower temperatures the behavior is increasingly fragile,[48-50] so much so that by some measures water near ~235 K is perhaps the most fragile liquid.[51] However, at even lower temperatures, water is expected to become a strong liquid.[8, 52, 53] The origin of this "fragile to strong" transition in water, which occurs at ~225 K, has been extensively debated.[8, 53-57] Within the LLCP hypothesis or the singularity free scenario, water's structure at ~zero pressure will be dominated by the HDL structural motif at high temperatures (e.g., above room temperature). However, as the temperature decreases, the fraction of the liquid exhibiting the LDL motif will increase. The LDL motif, which more closely resembles the tetrahedral bonding geometry that is found in other network glasses, is presumably responsible for the strong behavior found here (see **Fig. 6**). An Arrhenius temperature dependence at high and low temperature (for HDL-like and LDL-like, respectively) connected by a relatively broad transition is supported by recent experimental results,[35, 45, 55, 58-61] thermodynamic considerations,[52, 54, 57] and molecular dynamics simulations.[56]

The current results demonstrate that molecular translational diffusion on the scale of a few nanometers occurs in water for 120 – 144 K. The range of diffusion coefficients reported previously for these temperatures span a wide range from none observed (or inferred) to, for example, ~$10^{-17}$ m$^2$/s at 130 K.[12, 13, 45, 62, 63] Experiments using X-ray photon-correlation spectroscopy (XPCS) in LDA found $D_{tr} \sim 5 \times 10^{-18}$ m$^2$/s at 130 K.[62] That study, which monitored powdered HDA samples as they converted to LDA, was primarily concerned with whether the observed transition occurred between amorphous solid or liquid states. While it provided strong evidence for liquid-like diffusion in both HDA and LDA near their respective $T_g$'s, some aspects of the experiment probably made it difficult to obtain accurate values for the diffusion coefficient in LDA. Specifically, the diffusion coefficient for LDA was obtained at 130 K as the sample was converting from HDA to LDA over the course of $10^3$ s. During the measurement, the diffusion coefficient dropped from ~$5 \times 10^{-17}$ m$^2$/s (during the first 400 s) to ~$5 \times 10^{-18}$ m$^2$/s (during the last 200 s). Based on our diffusion measurements, we suspect that the duration of the XPCS experiments was too short to measure the diffusion in LDA at (metastable) equilibrium and were instead still probing the later stages of the HDA-LDA transition. The typical distance a molecule diffuses during a time, $t$, is $x_{diff} \sim \sqrt{6 D_{tr} t}$ (in 3 dimensions). $T_g$ is often taken to be the temperature at which the structural relaxation time for a liquid is $\tau_\alpha \sim 100$ s. Taking $D_{tr} = 5 \times 10^{-18}$ m²/s at 130 K and $t = 100$ s, which is the typical the structural relaxation time for a liquid at $T_g$, one finds $x_{diff} \sim 55$ nm – a distance which heuristically seems to be far larger than needed for equilibration. In contrast, we find $D_{tr} = 6.25 \times 10^{-21}$ m²/s at 136 K, which gives $x_{diff} \sim 2$ nm for $t = 100$ s. Note that if the diffusion coefficient was $5 \times 10^{-18}$ m²/s at 130 K in our experiments, then a film with 20 ML of D₂O in the middle of a film with a total coverage of 200 ML would mix on a characteristic timescale of ~4 s, instead of the ~9 × 10³ s that was observed (see **Table S1**).

In their H/D exchange experiments, Fisher and Devlin argued that molecular rotations could account for their results, while diffusion could not.[12] As mentioned in the introduction, Johari already discussed how diffusion might account for their observations.[17] We note that Fisher and Devlin's results do not rule out diffusion – instead they imply that rotations created isolated HOD on a timescale that was short compared to diffusion. Because the relative rates of translational and rotational diffusion in water are unknown for the temperatures of those experiments (between 115 and 122 K), whether faster rotations accounts for Fisher and Devlin's results remain an open question. Understanding the effects of rotations versus translations is probably also germane for explaining the intriguing isotope effects observed in the glass transition temperatures for LDA, HDA, and ice VI.[13] However, the observed correlations among the glass transition temperatures do not exclude the possibility of translational diffusion in LDA.

H/D isotope effects influence the structure and dynamics of water.[64] For example, the temperature of maximum density at ambient pressure for D₂O is 7.2 K higher than it is for H₂O. Also, diffusion is faster and viscosity is lower in H₂O compared to D₂O in normal and moderately supercooled water (e.g. 245 K to 300 K). Early work, which suggested a possible singularity in the H₂O's properties at ~128 K, found similar behavior in D₂O with an apparent singularity at ~133 K.[16] Isotope effects are also notable in amorphous ices at cryogenic temperatures (e.g., the ~4 K shift in $T_g$ between D₂O and H₂O discussed above).[13, 65] Furthermore, the dielectric relaxation times in D₂O are shifted by up to 12 K near $T_g$.[66] While most of experiments on isotope effects have focused on pure H₂O and D₂O, experiments using isotopic mixtures typically show that the property of interest (e.g., density or viscosity) vary smoothly between the 2 pure endpoints as the composition is changed.[67-69] Because the experiments reported here involve mixtures of H₂O and D₂O it is likely that the diffusion coefficients shown in **Fig. 6** are smaller (larger) than the coefficients for pure H₂O (D₂O). However, the magnitude of the isotope effect on diffusion at these low temperatures remains to be determined.

In a related set of experiments, some of us (RSS, BDK, and GAK) measured the diffusion in supercooled water by monitoring water desorption from nanoscale films of H₂¹⁶O and H₂¹⁸O.[70] Because isotope effects are typically smaller in H₂¹⁸O and H₂¹⁶O compared to those in H₂O and D₂O, those experiments should more closely reflect the diffusion coefficient in ordinary water. Overall, the agreement between the two distinct sets of experiments is quite good, and the differences between them are within expectations based on the isotope effects (see **Fig. S12**).

## V. CONCLUSIONS

The experiments reported here used IR spectroscopy to track the motion of intact $D_2O$ molecules within majority $H_2O$ films with coverages up to 300 ML, which corresponded to film thicknesses up to ~100 nm. Isotopically-layered films of $D_2O$ and $H_2O$ were grown on a Pt(111) surface at 108 K – a temperature at which diffusion was negligible – and subsequently annealed at temperatures from 120 to 144 K. The experiments were conducted in conditions where H/D exchange was minimized such that the results were insensitive to the rotational diffusion of the water molecules. In particular, if there was only rotational diffusion and no translational diffusion, the IR spectra would not have changed as the films were annealed. Instead, the results demonstrate that the initial, non-uniform distribution of $D_2O$ and $H_2O$ within the water films evolved toward a uniform distribution through translational diffusion of the intact water molecules. For experiments with the $D_2O$ probe layers initially sandwiched between 2 $H_2O$ layers, the rate of diffusive mixing was independent of the total thickness, $x_{film}$, for (at least) $x_{film} >$ 20 nm. For films with a total coverage of 200 ML, any variations of the diffusion coefficient versus the distance from the vacuum interface were below the uncertainty of the measurements (i.e. $< \pm$ 30%). The translational diffusion coefficient had an Arrhenius temperature dependence with an activation energy of 40.8 kJ/mol showing that water is strong liquid at and near the calorimetric glass transition at 136 K. The results also indicate that LDA, ASW, and HGW all relax to the same supercooled liquid water state upon sufficient annealing at and near $T_g$.

**SUPPLEMENTARY MATERIAL**

The supplementary material includes figures showing various IR spectra in support of discussion points in the main manuscript (**Figs. S1, S2, S3,** and **S8**). **Figs. S4** and **S9** show the IR integrals vs time for all the experiments used to determine the diffusion coefficients shown in **Fig 6**. Those figures also show the effect of increasing or decreasing the estimates of the characteristic times by ± 20% (Fig. S4) or ± 30% (Fig. S9) on the analysis. The characteristic times (see **Fig. S5**) and diffusion coefficients are also given in **Tables 1** and **2**. **Figures S10** and **S11** provide additional information for the results shown in Fig. 5. **Figure S6** shows IR integrals vs time for 200 ML water films with 5 ML $D_2O$ layer in the middle, and **Fig. S7** shows the same data in **Fig. 3** but with a linear time axis. Finally, **Fig. S12** compares the diffusion coefficients in the current work to those of a recent experiment.

ACKNOWLEDGEMENT

This work was supported by the U.S. Department of Energy (DOE), Office of Science, Office of Basic Energy Sciences, Division of Chemical Sciences, Geosciences, and Biosciences, Condensed Phase and Interfacial Molecular Science program, FWP 16248.

AUTHOR DECLARATIONS

**Conflict of Interest**

The authors have no conflict of interests to disclose.

**Author Contributions**

The work was conceptualized by GAK and BDK. GAK and MKD performed the experiments and analyzed the data. KG performed preliminary experiments. All authors contributed ideas to refine the experiments. GAK wrote the paper with input from all the authors.

DATA AVAILABILITY

The data that support the findings of this study are available within the article and its supplementary material.

REFERENCES:

[1] E. F. Burton, and W. F. Oliver, "X-ray diffraction patterns of ice," Nature **135,** 505-506 (1935).

[2] E. Mayer, "New methods for vitrifying water and other liquids by rapid cooling of their aerosols," J. App. Phys. **58,** 663 (1985).

[3] O. Mishima, L. D. Calvert, and E. Whalley, "Melting ice I at 77 K and 10 kbar - A new method of making amorphous solids," Nature **310,** 393-395 (1984).

[4] O. Mishima, L. D. Calvert, and E. Whalley, "An apparently 1st-order transition between 2 amorphous phases of ice induced by pressure," Nature **314,** 76-78 (1985).

[5] T. Loerting, C. Salzmann, I. Kohl, E. Mayer, and A. Hallbrucker, "A second distinct structural "state" of high-density amorphous ice at 77 K and 1 bar," Phys. Chem. Chem. Phys. **3,** 5355-5357 (2001).

[6] A. Rosu-Finsen, M. B. Davies, A. Amon, H. Wu, A. Sella, A. Michaelides, and C. G. Salzmann, "Medium-density amorphous ice," Science **379,** 474-478 (2023).

[7] O. Mishima, and H. E. Stanley, "The relationship between liquid, supercooled and glassy water," Nature **396,** 329-335 (1998).

[8] C. A. Angell, "Insights into phases of liquid water from study of its unusual glass-forming properties," Science **319**, 582-587 (2008).

[9] K. Amann-Winkel, R. Bohmer, F. Fujara, C. Gainaru, B. Geil, and T. Loerting, "Colloquium: Water's controversial glass transitions," Rev. Mod. Phys. **88**, 011002 (2016).

[10] A. Hallbrucker, E. Mayer, and G. P. Johari, "Glass-liquid transition and the enthalpy of devitrification of annealed vapor deposited ASW," J. Phys. Chem. **93**, 4986 (1989).

[11] A. Hallbrucker, E. Mayer, and G. P. Johari, "The heat Capacity and glass transition of hyperquenched glassy water," Philos. Mag. B. **60**, 179 (1989).

[12] M. Fisher, and J. P. Devlin, "Defect activity in amorphous ice from isotopic exchange data: Insight into the glass transition," J. Phys. Chem. **99**, 11584 (1995).

[13] J. J. Shephard, and C. G. Salzmann, "Molecular reorientation dynamics govern the glass transitions of the amorphous ices," J. Phys. Chem. Lett. **7**, 2281-2285 (2016).

[14] P. G. Debenedetti, "Supercooled and glassy water," J. Phys. Condens. Matter **15**, R1669-R1726 (2003).

[15] P. Gallo, K. Arnann-Winkel, C. A. Angell, M. A. Anisimov, F. Caupin, C. Chakravarty, E. Lascaris, T. Loerting, A. Z. Panagiotopoulos, J. Russo, J. A. Sellberg, H. E. Stanley, H. Tanaka, C. Vega, L. M. Xu, and L. G. M. Pettersson, "Water: A tale of two liquids," Chem. Rev. **116**, 7463-7500 (2016).

[16] C. A. Angell, "Supercooled water," Ann. Rev. Phys. Chem. **34**, 593-630 (1983).

[17] G. P. Johari, "Amorphous solid water's isotopic exchange kinetics," J. Chem. Phys. **117**, 2782-2789 (2002).

[18] R. S. Smith, N. G. Petrik, G. A. Kimmel, and B. D. Kay, "Thermal and nonthermal physiochemical processes in nanoscale films of amorphous solid water," Acc. Chem. Res. **45**, 33-42 (2012).

[19] R. S. Smith, and B. D. Kay, "The existence of supercooled liquid water at 150 K," Nature **398**, 788-791 (1999).

[20] R. S. Smith, Z. Dohnálek, G. A. Kimmel, K. P. Stevenson, and Bruce D. Kay, " Chem. Phys., 258, 291 (2000). "The self-diffusivity of amorphous solid water near 150 K," Chem. Phys. **258**, 291-305 (2000).

[21] S. M. McClure, E. T. Barlow, M. C. Akin, D. J. Safarik, T. M. Truskett, and C. B. Mullins, "Transport in amorphous solid water films: Implications for self-diffusivity," J. Phys. Chem. B **110**, 17987-17997 (2006).

[22] S. M. McClure, D. J. Safarik, T. M. Truskett, and C. B. Mullins, "Evidence that amorphous water below 160 K is not a fragile liquid," J. Phys. Chem. B **110**, 11033-11036 (2006).

[23] R. S. Smith, N. G. Petrik, G. A. Kimmel, and B. D. Kay, "Communication: Proton exchange in low temperature co-mixed amorphous $H_2O$ and $D_2O$ films: The effect of the underlying Pt(111) and graphene substrates," J. Chem. Phys. **149**, 081104 (2018).

[24] J. W. Cahn, W. B. Hillig, and G. W. Sears, "Molecular mechanism of solidification," Acta Metall. **12,** 1421-1439 (1964).

[25] J. Q. Broughton, G. H. Gilmer, and K. A. Jackson, "Crystallization rates of a Lennard-Jones liquid," Phys. Rev. Lett. **49,** 1496-1500 (1982).

[26] M. L. Ferreira Nascimento, and E. D. Zanotto, "Does viscosity describe the kinetic barrier for crystal growth from the liquidus to the glass transition?," J. Chem. Phys. **133,** 174701 (2010).

[27] M. K. Dunlap, L. Kringle, B. D. Kay, and G. A. Kimmel, "Proton diffusion and hydrogen/deuterium exchange in amorphous solid water at temperatures from 114 to 134 K," J. Chem. Phys. **161,** 244504 (2024).

[28] G. A. Kimmel, K. P. Stevenson, Z. Dohnálek, R. S. Smith, and B. D. Kay, "Control of amorphous solid water morphology using molecular beams - I: Experimental results," J. Chem. Phys. **114,** 5284 (2001).

[29] K. P. Stevenson, G. A. Kimmel, Z. Dohnálek, R. S. Smith, and B. D. Kay, "Controlling the morphology of amorphous solid water," Science **283,** 1505 (1999).

[30] S. Nie, P. J. Feibelman, N. C. Bartelt, and K. Thurmer, "Pentagons and heptagons in the first water layer on Pt(111)," Phys. Rev. Lett. **105,** 026102 (2010).

[31] T. Loerting, M. Bauer, I. Kohl, K. Watschinger, K. Winkel, and E. Mayer, "Cryoflotation: Densities of amorphous and crystalline ices," J. Phys. Chem. B **115,** 14167-14175 (2011).

[32] C. Q. Yuan, R. S. Smith, and B. D. Kay, "Surface and bulk crystallization of amorphous solid water films: Confirmation of "top-down" crystallization," Surf. Sci. **652,** 350-354 (2016).

[33] M. D. Ediger, "Perspective: Highly stable vapor-deposited glasses," J. Chem. Phys. **147,** 210901 (2017).

[34] S. F. Swallen, K. L. Kearns, M. K. Mapes, Y. S. Kim, R. J. McMahon, M. D. Ediger, T. Wu, L. Yu, and S. Satija, "Organic glasses with exceptional thermodynamic and kinetic stability," Science **315,** 353-356 (2007).

[35] L. Kringle, W. A. Thornley, B. D. Kay, and G. A. Kimmel, "Reversible structural transformations in supercooled liquid water from 135 to 245 K," Science **369,** 1490-1492 (2020).

[36] L. Kringle, W. A. Thornley, B. D. Kay, and G. A. Kimmel, "Structural relaxation and crystallization in supercooled water from 170 to 260 K," Proc. Natl. Acad. Sci. U. S. A. **118,** e2022884118 e2022884118 (2021).

[37] H. J. Bakker, and J. L. Skinner, "Vibrational spectroscopy as a probe of structure and dynamics in liquid water," Chem. Rev. **110,** 1498-1517 (2010).

[38] G. Cicero, J. C. Grossman, E. Schwegler, F. Gygi, and G. Galli, "Water confined in nanotubes and between graphene sheets: A first principle study," J. Am. Chem. Soc. **130,** 1871-1878 (2008).

[39] D. T. Limmer, A. P. Willard, P. Madden, and D. Chandler, "Hydration of metal surfaces can be dynamically heterogeneous and hydrophobic," Proc. Natl. Acad. Sci. U. S. A. **110,** 4200-4205 (2013).

[40] A. P. Willard, and D. Chandler, "The molecular structure of the interface between water and a hydrophobic substrate is liquid-vapor like," J. Chem. Phys. **141,** 18c519 (2014).

[41] B. Rowland, and J. P. Devlin, "Spectra of dangling OH groups at ice cluster surfaces and within pores of amorphous ice," J. Chem. Phys. **94,** 812 (1991).

[42] V. Buch, and J. P. Devlin, "Spectra of dangling OH bonds in amorphous ice: Assignment to 2- 3-coordinated surface molecules," J. Chem. Phys. **94,** 4091 (1991).

[43] M. K. Mapes, S. F. Swallen, and M. D. Ediger, "Self-diffusion of supercooled o-terphenyl near the glass transition temperature," J. Phys. Chem. B **110,** 507-511 (2006).

[44] K. L. Ngai, J. H. Magill, and D. J. Plazek, "Flow, diffusion and crystallization of supercooled liquids: Revisited," J. Chem. Phys. **112,** 1887-1892 (2000).

[45] Y. T. Xu, N. G. Petrik, R. S. Smith, B. D. Kay, and G. A. Kimmel, "Growth rate of crystalline ice and the diffusivity of supercooled water from 126 to 262 K," Proc. Natl. Acad. Sci. U. S. A. **113,** 14921-14925 (2016).

[46] L. Berthier, and G. Biroli, "Theoretical perspective on the glass transition and amorphous materials," Rev. Mod. Phys. **83,** 587-645 (2011).

[47] M. D. Ediger, "Spatially heterogeneous dynamics in supercooled liquids," Ann. Rev. Phys. Chem. **51,** 99-128 (2000).

[48] W. S. Price, H. Ide, and Y. Arata, "Self-diffusion of supercooled water to 238 K using PGSE NMR diffusion measurements," J. Phys. Chem. A **103,** 448-450 (1999).

[49] A. Dehaoui, B. Issenmann, and F. Caupin, "Viscosity of deeply supercooled water and its coupling to molecular diffusion," Proc. Natl. Acad. Sci. U. S. A. **112,** 12020-12025 (2015).

[50] L. P. Singh, B. Issenmann, and F. Caupin, "Pressure dependence of viscosity in supercooled water and a unified approach for thermodynamic and dynamic anomalies of water," Proc. Natl. Acad. Sci. U. S. A. **114,** 4312-4317 (2017).

[51] K. Ito, C. T. Moynihan, and C. A. Angell, "Thermodynamic determination of fragility in liquids and a fragile-to-strong liquid transition in water," Nature **398,** 492-495 (1999).

[52] C. A. Angell, C. T. Moynihan, and M. Hemmati, "'Strong' and 'superstrong' liquids, and an approach to the perfect glass state via phase transition," J. Non-Cryst. Solids **274,** 319-331 (2000).

[53] M. De Marzio, G. Camisasca, M. Rovere, and P. Gallo, "Fragile to strong crossover and Widom line in supercooled water: A comparative study," Front. Phys. **13,** 136103 (2018).

[54] F. W. Starr, C. A. Angell, and H. E. Stanley, "Prediction of entropy and dynamic properties of water below the homogeneous nucleation temperature," Physica A **323,** 51-66 (2003).

[55] N. J. Hestand, and J. L. Skinner, "Perspective: Crossing the Widom line in no man's land: Experiments, simulations, and the location of the liquid-liquid critical point in supercooled water," J. Chem. Phys. **149,** 140901 (2018).

[56] R. Shi, J. Russo, and H. Tanaka, "Origin of the emergent fragile-to-strong transition in supercooled water," Proc. Natl. Acad. Sci. U. S. A. **115,** 9444-9449 (2018).

[57] F. Caupin, "Predictions for the properties of water below its homogeneous crystallization temperature revisited," J. Non-Cryst. Solids: X **14,** 100090 (2022).

[58] J. A. Sellberg, C. Huang, T. A. McQueen, N. D. Loh, H. Laksmono, D. Schlesinger, R. G. Sierra, D. Nordlund, C. Y. Hampton, D. Starodub, D. P. DePonte, M. Beye, C. Chen, A. V. Martin, A. Barty, K. T. Wikfeldt, T. M. Weiss, C. Caronna, J. Feldkamp, L. B. Skinner, M. M. Seibert, M. Messerschmidt, G. J. Williams, S. Boutet, L. G. M. Pettersson, M. J. Bogan, and A. Nilsson, "Ultrafast X-ray probing of water structure below the homogeneous ice nucleation temperature," Nature **510,** 381-384 (2014).

[59] K. H. Kim, A. Spah, H. Pathak, F. Perakis, D. Mariedahl, K. Amann-Winkel, J. A. Sellberg, J. H. Lee, S. Kim, J. Park, K. H. Nam, T. Katayama, and A. Nilsson, "Maxima in the thermodynamic response and correlation functions of deeply supercooled water," Science **358,** 1589-1593 (2017).

[60] C. R. Krueger, N. J. Mowry, G. Bongiovanni, M. Drabbels, and U. J. Lorenz, "Electron diffraction of deeply supercooled water in no man's land," Nat. Commun. **14,** 2812 4 (2023).

[61] L. Kringle, B. D. Kay, and G. A. Kimmel, "Structural relaxation of water during rapid cooling from ambient temperatures," J. Chem. Phys. **159,** 064509 (2023).

[62] F. Perakis, K. Amann-Winkel, F. Lehmkuhler, M. Sprung, D. Mariedahl, J. A. Sellberg, H. Pathak, A. Spah, F. Cavalca, D. Schlesinger, A. Ricci, A. Jain, B. Massani, F. Aubree, C. J. Benmore, T. Loerting, G. Grubel, L. G. M. Pettersson, and A. Nilsson, "Diffusive dynamics during the high-to-low density transition in amorphous ice," Proc. Natl. Acad. Sci. U. S. A. **114,** 8193-8198 (2017).

[63] P. Ghesquière, T. Mineva, D. Talbi, P. Theulé, J. A. Noble, and T. Chiavassa, "Diffusion of molecules in the bulk of a low density amorphous ice from molecular dynamics simulations," Phys. Chem. Chem. Phys. **17,** 11455-11468 (2015).

[64] M. Ceriotti, W. Fang, P. G. Kusalik, R. H. McKenzie, A. Michaelides, M. A. Morales, and T. E. Markland, "Nuclear quantum effects in water and aqueous systems: Experiment, theory, and current challenges," Chem. Rev. **116,** 7529-7550 (2016).

[65] G. P. Johari, A. Hallbrucker, and E. Mayer, "Isotope and impurity effects on the glass-transition and crystallization of pressure-amorphized hexagonal and cubic ice," J. Chem. Phys. **95,** 6849-6855 (1991).

[66] C. Gainaru, A. L. Agapov, V. Fuentes-Landete, K. Amann-Winkel, H. Nelson, K. W. Koster, A. I. Kolesnikov, V. N. Novikov, R. Richert, R. Bohmer, T. Loerting, and A. P. Sokolov, "Anomalously large isotope effect in the glass transition of water," Proc. Natl. Acad. Sci. U. S. A. **111,** 17402-17407 (2014).

[67] E. Swift, "The temperature of maximum density of $D_2O$ and of its mixtures with $H_2O$," J. Am. Chem. Soc. **61,** 1293-1294 (1939).

[68] U. Kaatze, "Dielectric-relaxation of $H_2O/D_2O$ mixtures," Chem. Phys. Lett. **203,** 1-4 (1993).

[69] O. Gajst, R. Simkovitch, and D. Huppert, "Anomalous $H^+$ and $D^+$ excited-state proton-transfer rate in $H_2O/D_2O$ mixtures," J. Phys. Chem. A **121,** 6917-6924 (2017).

[70] R. S. Smith, W. A. Thornley, G. A. Kimmel, and B. D. Kay, "Supercooled liquid Water diffusivity at temperatures near the glass transition temperature," arXiv:2502.02876 [cond-mat.soft] (2025).

Supplementary Material for

**Translational diffusion in supercooled water at and near the glass transition temperature – 136 K**

Greg A. Kimmel,* Megan K. Dunlap, Kirill Gurdumov, R. Scott Smith, Loni Kringle, and Bruce D. Kay

**Section 1. Time-dependence of diffusive mixing for water films with the same geometry annealed at different temperatures.**

In diffusion problems, a natural unit of time, $t'$, is given by the diffusion coefficient and a typical length scale, $\lambda$, for the system of interest: $t' = D_{tr}(T)t/\lambda^2 = t/\tau(T)$, and a natural unit of length is $x' = x/\lambda$. For example, for a substance that is initially localized in a layer from $-\lambda < x < \lambda$ in an infinite film, the concentration as a function of space and time in the reduced units, $C(x', t')$, has a particularly simple form:[1]

$$C(x', t') = \frac{1}{2}C_0\left\{\text{erf}\left(\frac{1-x'}{2\sqrt{t'}}\right) + \text{erf}\left(\frac{1+x'}{2\sqrt{t'}}\right)\right\}. \tag{S1}$$

For experiments such as those shown in Figs. 3 and S4, this analytical solution will be an excellent approximation for the concentration (but not the IR integrals!) until the concentration of $D_2O$ at the water/Pt and water/vacuum interfaces becomes appreciable. It is clear from Eqn. S1, that there should be a characteristic time for each annealing temperature, $\tau(T) = \lambda^2/D_{tr}(T)$ that gives the similarity scaling observed in Figs. 3b and S4b. To determine $\tau(T)$ for all the temperatures, we started by fixing its value at the glass transition temperature: $\tau(T_g) = \lambda^2/D_{tr}(T_g)$. Following Crank[1], we chose $\lambda$ as one half the width of the 20 ML $D_2O$ layer: $\lambda = 3.3$ nm. For the diffusion rate at $T_g$, we used the results from the experiments measuring the dangling OD signal at the vacuum interface (see Figs. 4, 5, and 6): $D_{tr}(T_g) = 6.25 \times 10^{-21}$ m²/s, which gives $\tau(T_g) = 1742$ s. This scale factor was then applied to one of the data sets take at 136 K. Values of $\tau(T)$ were iteratively optimized for all the other data sets to produce the lowest overall scatter. The results of that process are shown in Fig. S4, and Table S1. Note that any arbitrary choice for $\tau(T_g)$, e.g. $\tau(T_g) = 1$ s (i.e., no rescaling of the initial data set at 136 K) would not have changed the ultimate determination of $D_{tr}(T)$ shown in Fig. 6 and Table S1. Instead, it would have led to a rescaling of the $\tau(T)$ values shown in Table 1 by a constant factor. Another way to state this issue is that the similarity scaling shown in Figs. 3b and S4b allows us to uniquely determine the ratio of characteristic times at any two temperatures, while the absolute values of $\tau(T)$ can only be determined to within common constant factor.

For the experiments shown in Figs 4 and 5, $D_{tr}(T)$ was determined by comparing the dangling OD signal vs time to the results of diffusion calculations (as described in the main text). The characteristic times, $\tau(T)$, which were used to display the similarity scaling for those experiments (see Fig. S9, and Table S2), were determined only to within a single common constant factor (following the same arguments described above). We took advantage of this flexibility to choose the overall scale factor such that the characteristic times for the experiments with a $D_2O$ layer at and near the surface approximately match to those with 20 ML $D_2O$ in the middle of 180 ML $H_2O$ layer. With this choice for the scaling, the good agreement of the temperature dependence for the two sets of experiments is readily apparent (see Fig. S5).

**Supplementary Figures**

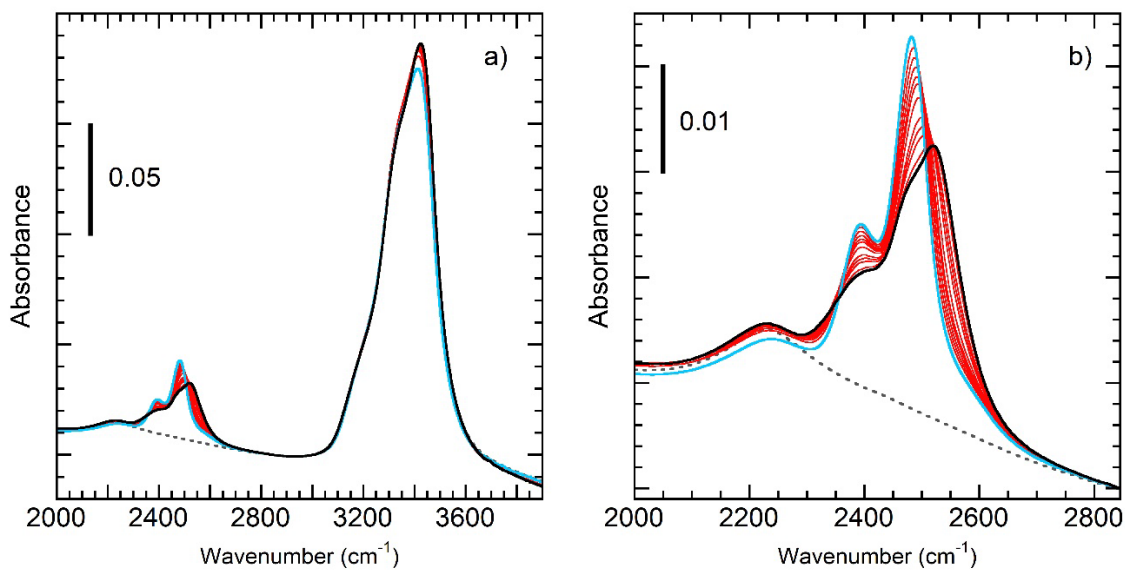

**Fig. S1**. Representative IRAS spectra for a film with a 20 ML $D_2O$ layer in the middle 180 ML $H_2O$ deposited on Pt(111) and then annealed at 132.5 K. a) Spectra showing both the OH-stretch (3000 – 3700 cm$^{-1}$) and OD-stretch regions (~2250 – 2750 cm$^{-1}$). b) The same spectra focusing on the OD-stretch region. The solid black line shows the first spectrum obtained at 132.5 K. The blue line shows the spectrum after annealing for 12,430 s. The red lines show spectra at $t$ = 82, 164, 370, 660, 1080, 2650, 4510, 6020, 9040, and $1.38 \times 10^4$ s. The IR spectrum for a 180 ML $H_2O$ film without any $D_2O$ (black dotted line) has a broad "association band," attributed to the $H_2O$ bending mode plus librations,[2, 3] that overlaps with the OD-stretching region. For the results shown in **Fig. 2a**, the spectrum for 180 ML $H_2O$ has been subtracted from all the spectra.

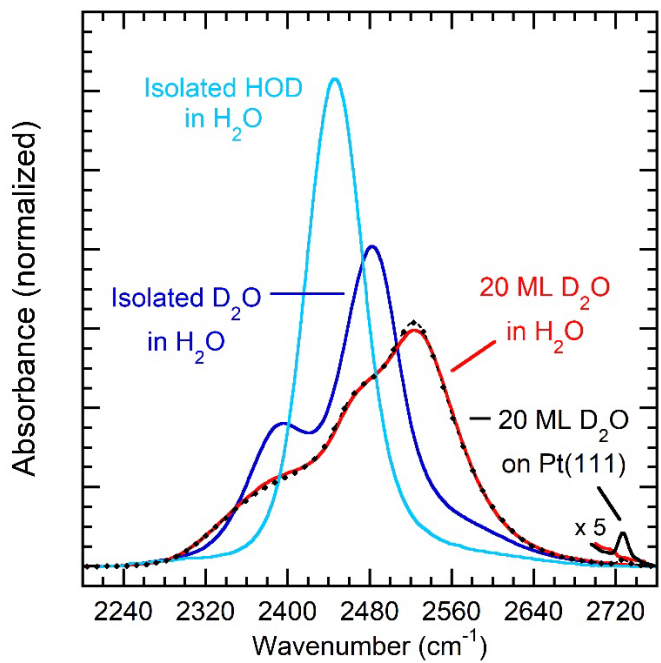

**Fig. S2**. Infrared spectra in the OD-stretch region for water films with a total coverage of 200 ML with ~10% isolated HOD in $H_2O$ (light blue line), and 10% isolated $D_2O$ in $H_2O$ (dark blue line), and 20 ML $D_2O$ in a single layer embedded in the middle of the $H_2O$ (red line). To account for optical effects associated with the different film thickness and differences in the total amounts of D atoms in the films, the spectra have been normalized by their respective integrals over the OD-stretch band (i.e., 2230 – 2760 cm$^{-1}$). The spectrum for 20 ML $D_2O$ on Pt(111) is shown for comparison (black line and symbols). Its peak at 2727 cm$^{-1}$ is due to $D_2O$ molecules with a single non-bonded OD group (a "dangling OD") at the vacuum interface, which is absent for the $D_2O$ layer in the middle of the $H_2O$.

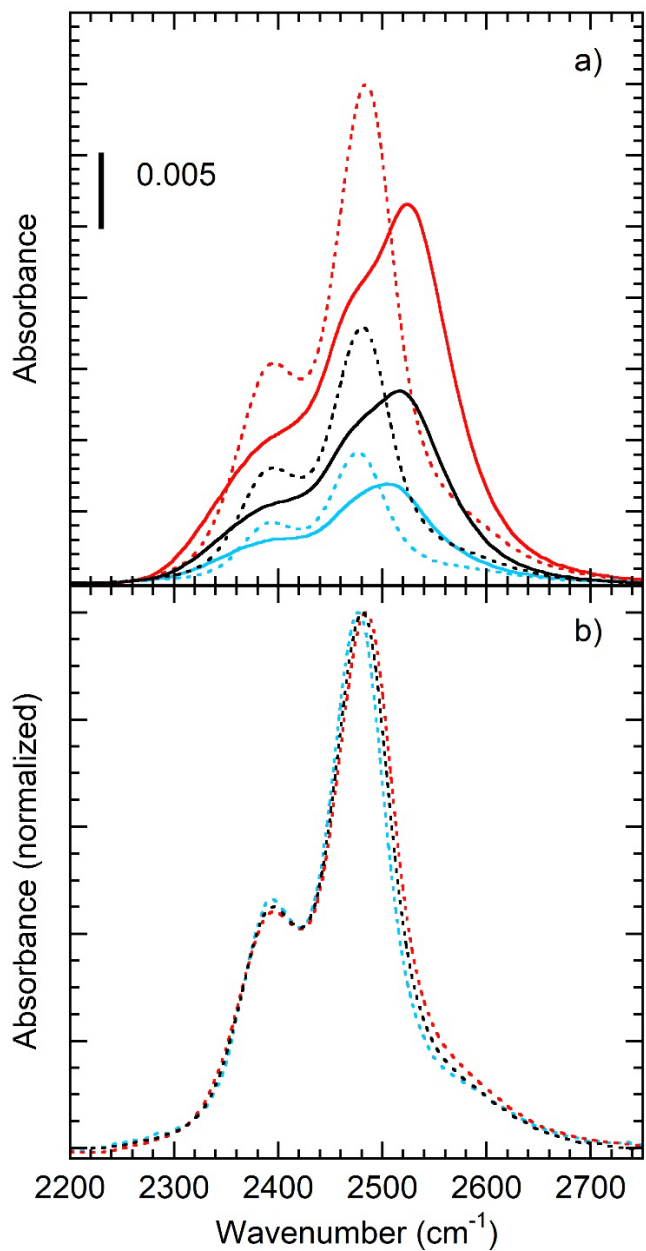

**Fig. S3.** IR spectra in the OD-stretch region for water films with 5 ML D$_2$O + 195 ML H$_2$O (blue), 10 ML D$_2$O + 190 ML H$_2$O (black), and 20 ML D$_2$O + 180 ML H$_2$O (red). The D$_2$O layers were deposited in the middle of the corresponding H$_2$O layers. a) The as-grown spectra (solid lines) reflect the differences in the initial configurations. The absorbance is also larger for the thicker D$_2$O layers. The dotted lines show the spectra obtained after annealing at 136 K. b) The IR spectra after annealing that have been normalized to the peak intensity. Independent of the different starting configurations, the lineshapes of all the final spectra are similar and representative of isolated D$_2$O in H$_2$O. (In contrast, the initial spectra are all distinct, as shown, for example in **Fig. 1b**.)

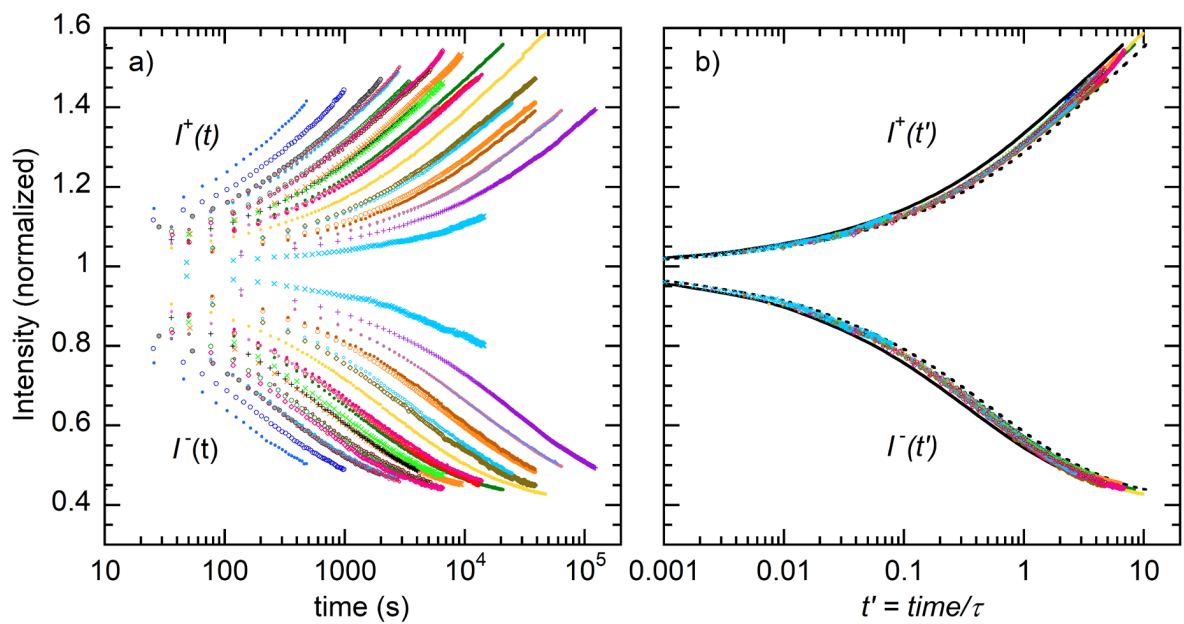

**Fig. S4.** (a) Integrals over low wavenumber (2365 – 2495 cm$^{-1}$, $I^+(t)$) and high wavenumber (2525 – 2670 cm$^{-1}$, $I^-(t)$) regions of the OD-stretch band vs time for water films with a 20 ML D$_2$O layer in the middle of 180 ML H$_2$O. The results for 24 experiments (including those shown in **Fig. 3**) are shown. Each symbol in the figure corresponds to an IR spectrum taken at the corresponding temperature. **Table S1** lists the annealing temperatures and characteristic times, $\tau$, for all the experiments. $I^+(t)$ is associated with the increasing fraction of isolated D$_2$O as the films anneal, while $I^-(t)$ reflects the loss of "bulk-like" D$_2$O. b) When the times in a) are scaled by $\tau$, all the data collapse onto 2 curves (i.e., $I^+(t')$ and $I^-(t')$ where $t' = t/\tau$) as expected for diffusion. To assess the uncertainty in determining $\tau$, the averages of $I^+(t')$ and $I^-(t')$ were calculated: $I_{ave}^+(t')$ and $I_{ave}^-(t')$, respectively. The dotted black lines show the results of increasing the characteristic times by 20% (i.e., $I_{ave}^+(t/1.2\tau)$, and $I_{ave}^-(t/1.2\tau)$), while the dotted black show the result of decreasing $\tau$ by 20%. The results suggest that $\tau$ can be determined to within ± 20% in these experiments.

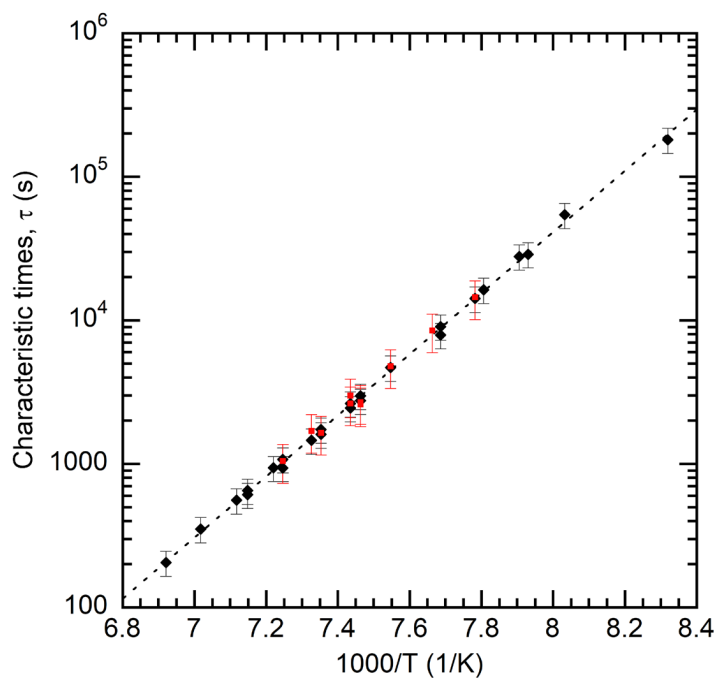

**Fig. S5.** Characteristic times, $\tau$, versus 1000/T for diffusive mixing of layered $D_2O/H_2O$ films. The black diamonds show $\tau$ for films with 20 ML $D_2O$ in the middle of 180 ML $H_2O$ (see **Fig. S4.**) The dotted line shows an Arrhenius fit to the data with an activation energy of 40.76 kJ/mol. (As discussed in **Section S1**, the values of $\tau(T)$ are determined only to within an overall scale factor, the magnitude of the prefactor is arbitrary.) The error bars show the estimated uncertainty as ± 20%. **Table S1** lists the temperatures characteristic times for these experiments. The red squares show $\tau$ for films with 20 ML $D_2O$ adsorbed on top of 180 ML $H_2O$ (see **Figs. 4** and **S9**), with an estimated uncertainty of ± 30%.

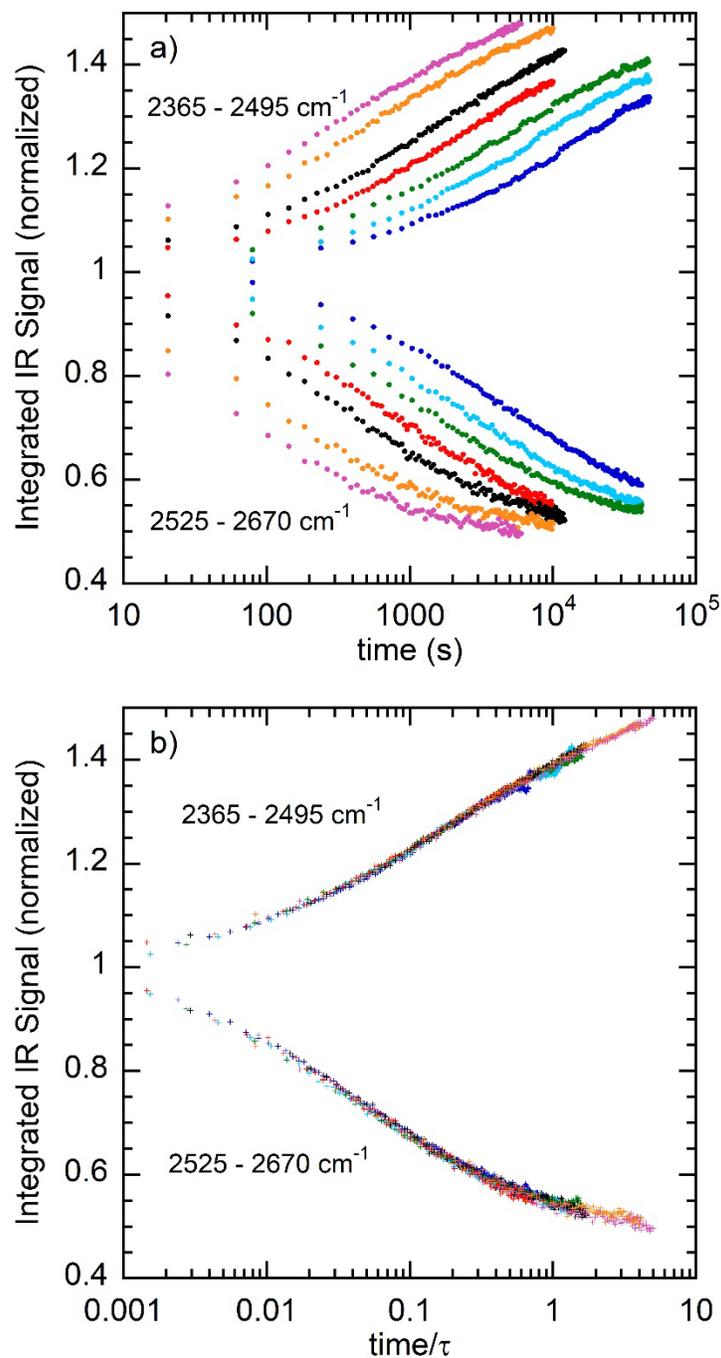

**Fig. S6.** Integrated IR signals versus time (a) and time/$\tau$ (b) for films that had 5 ML D$_2$O deposited in the middle of 195 ML H$_2$O. The integral over 2365 – 2495 cm$^{-1}$ reflects the increasing amount of isolated D$_2$O in the films, while the integral over 2525 – 2670 cm$^{-1}$ reflects the loss of "bulk-like" D$_2$O (see **Fig. 3**.) The films were annealed at 120.5 (dark blue), 122.5 (light blue), 124.5 (green), 126.5 (red), 128.5 (black), 132.5 (orange), and 134.5 K (purple). The characteristic times increase by a factor of ~83 over this temperature range. b) When normalized by a characteristic time, all the data collapse onto two common curves.

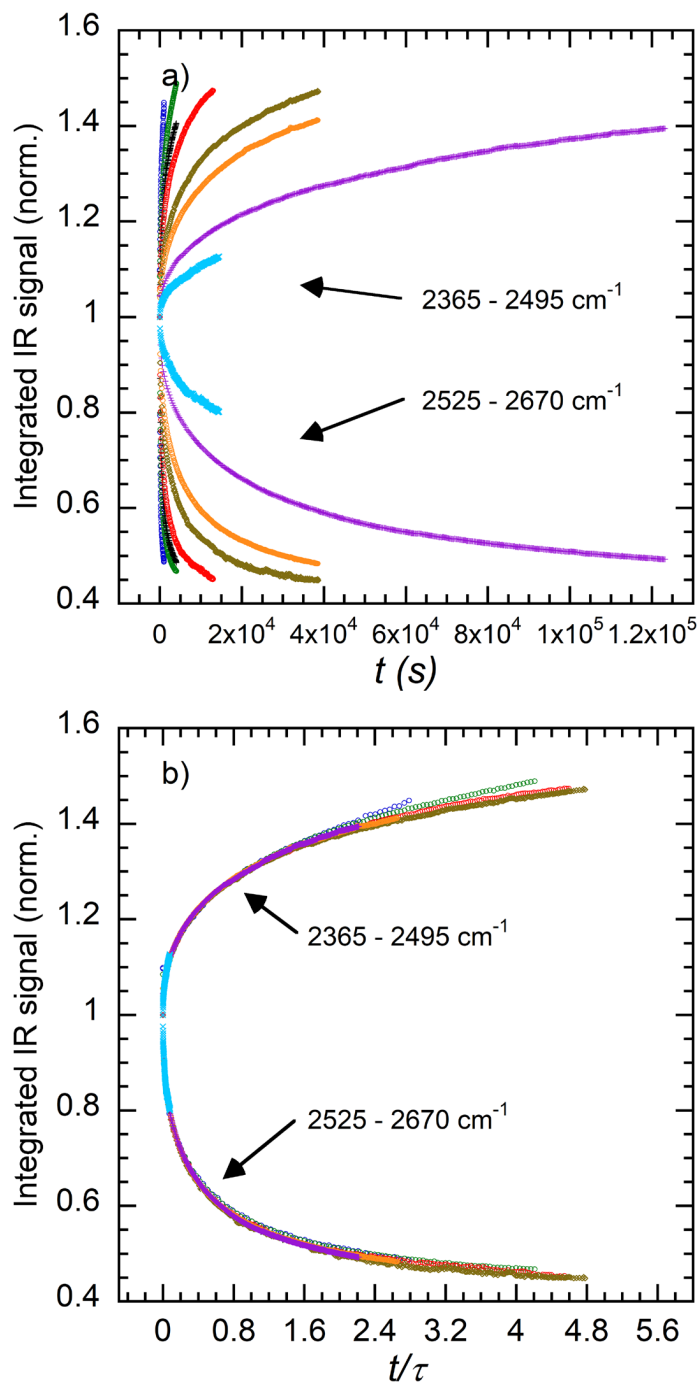

**Fig. S7.** Integrated IR signals of the OD-stretch band vs time. a) Integrals over low wavenumber (2365 – 2495 cm$^{-1}$) and high wavenumber (2525 – 2670 cm$^{-1}$) portions of the OD-stretch band. The increasing intensity of the signal in the low wavenumber portion of the band is associated with increasingly isolated D$_2$O, while the decreasing intensity at higher wavenumbers is due to the loss of "bulk-like" D$_2$O. b) When the data in a) is normalized by a characteristic time, $\tau$, all the data collapse onto 2 common curves, as expected for diffusion. **Fig. 4** displays this data using a logarithmic scale for the time.

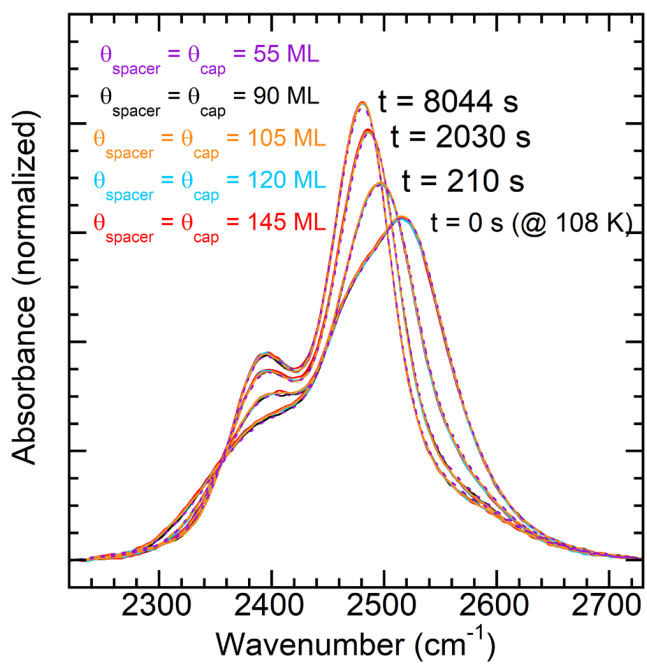

**Fig. S8.** IR spectra of water films annealed at 134 K. The films all had a 10 ML $D_2O$ layer deposited between 2 $H_2O$ layers with coverages ($\theta_{cap} = \theta_{spacer}$) of 55 (dashed purple), 90 (black), 105 (orange), 120 (blue) and 145 ML (red). After accounting for trivial changes in absorbance with increasing film thickness, all the spectra are essentially identical indicating that the diffusion is independent of film thickness for ~ 40 to 100 nm. Similar results (not shown) indicate that the diffusion is also independent of thickness for films as thin as ~15 nm.

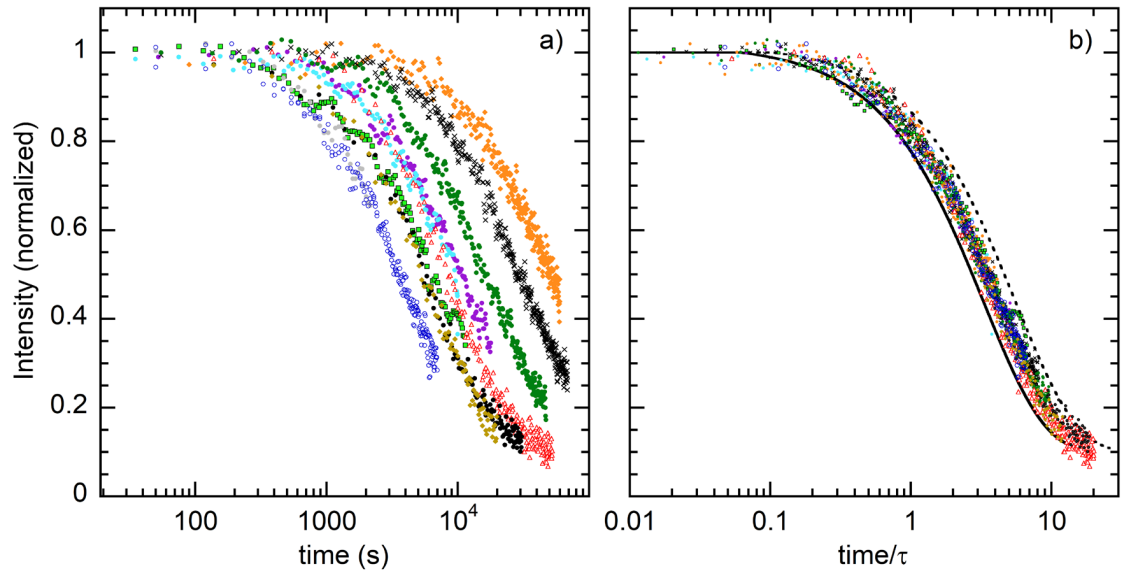

**Fig. S9.** a) The normalized integrated intensity of the dangling OD peak (symbols) versus time, $I^{dOD}(t)$, for water films with 20 ML D$_2$O deposited on 180 ML H$_2$O and subsequently annealed at 128.5 K (orange diamonds), 130.5 K (black crosses), 132.5 K (green circles), 134 K (red triangles), 134.5 K (purple and light blue circles), 136 K (olive diamonds and black circles), and 138 K (dark blue and grey circles). b) When normalized by a characteristic time, $\tau$, (see **Fig. S5)**, all the data collapse onto a common curve $I^{dOD}(t')$, where $t' = t/\tau$. To assess the uncertainty in determining $\tau$, the average of $I^{dOD}(t')$, $I^{dOD}_{ave}(t')$, was calculated for all the data sets. The solid [dotted] black lines show $I^{dOD}_{ave}(1.3t')$ [$I^{dOD}_{ave}(t'/1.3)$]. The results indicate that $\tau$ could be determined to within ± 30% for these experiments.

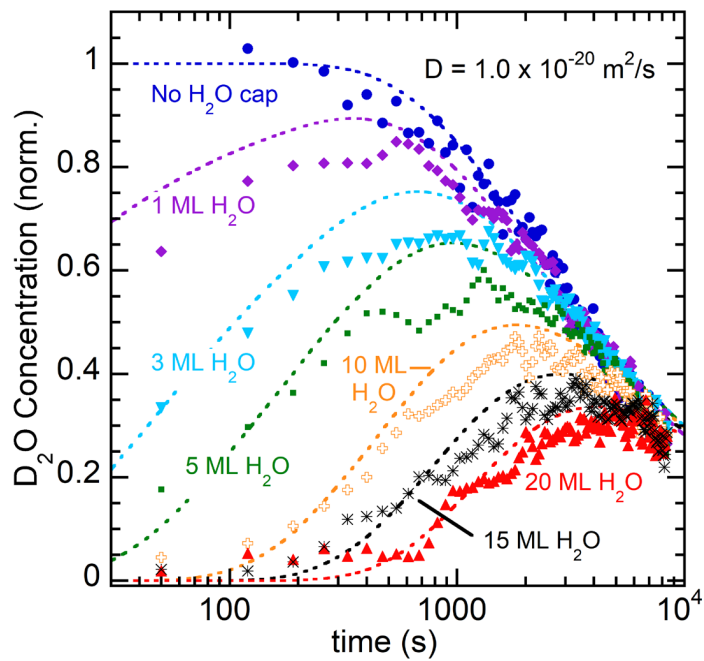

**Fig. S10.** a) The (normalized) concentration of $D_2O$ at the vacuum interface versus time for water films with 20 ML $D_2O$ and 180 ML $H_2O$ annealed at 138 K. The concentration was determined from the magnitude of the dangling OD peak in IR (symbols) and calculated assuming diffusive mixing of the layer water films (dotted lines). The calculations also account for water desorption during the annealing. When the $D_2O$ layer was deposited on top of a 180 ML film (blue circles, "No $H_2O$ cap"), the initial concentration was unity and subsequently decayed. However, when $H_2O$ layers of increasing thickness were adsorbed on top of the $D_2O$, the dangling OD signal initially increased as diffusion brought some $D_2O$ to the vacuum interface, but then the signal eventually decreased as diffusion continued to disperse the $D_2O$ into the bulk of the $H_2O$ film. The calculated signals (dotted lines) were ~3 to 10% larger than the observed signals. The IR spectra in these experiments indicated some H/D exchange at the vacuum interface led to small amounts of isolated HOD. This process, which was negligible for $D_2O$ layers in the middle of $H_2O$ films, was not included in the calculations. Future experiments are planned to investigate this effect in more detail. However, we believe it did not significantly impact the measured diffusion coefficient.

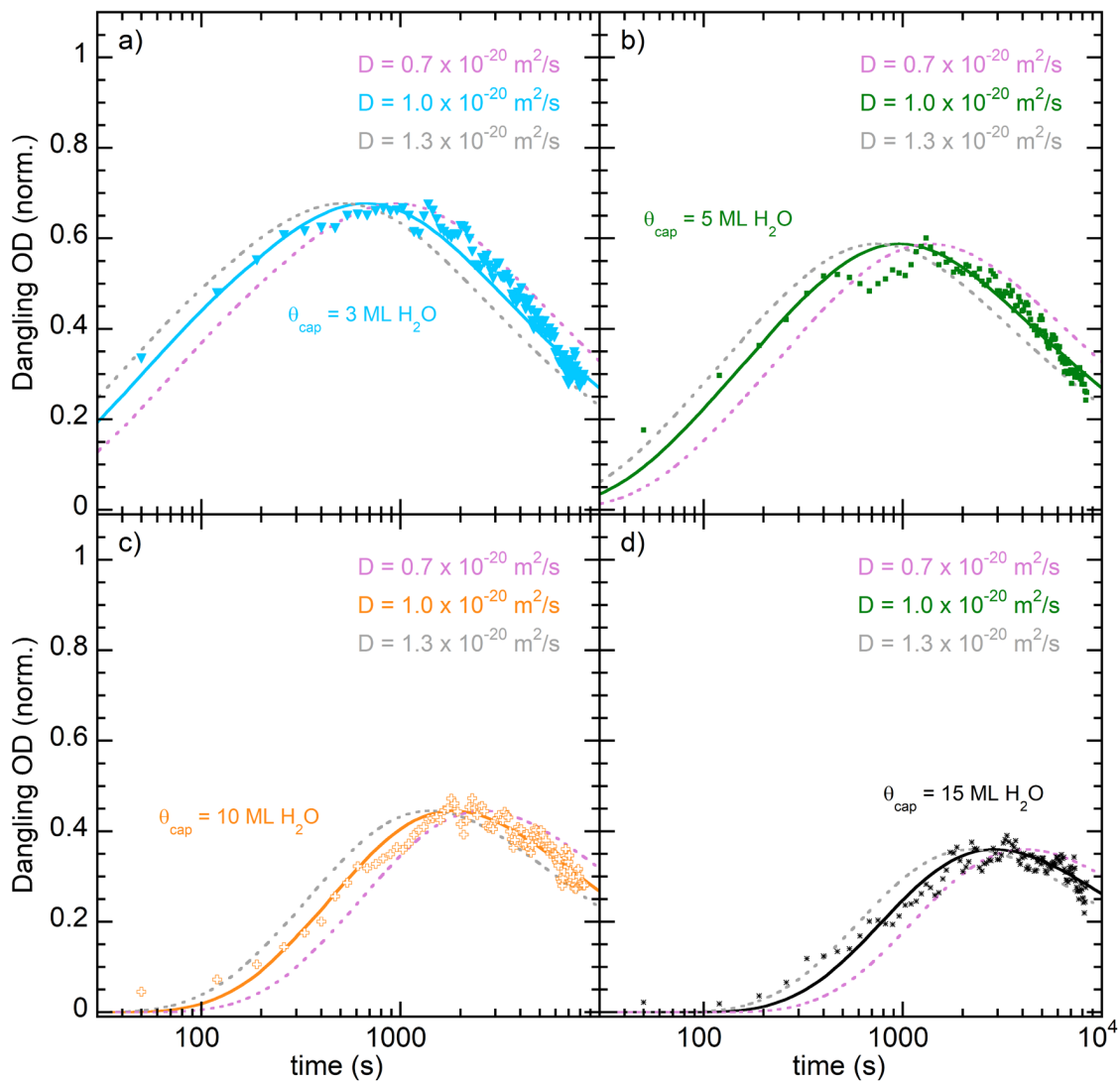

**Fig. S11.** This figure shows the how changing the calculated diffusion coefficient by ± 30% affects the calculation of the surface concentration versus time for experiments with a 20 ML D$_2$O layer capped by a) 3 ML H$_2$O, b) 5 ML H$_2$O, c) 10 ML H$_2$O, and d) 15 ML H$_2$O. The green lines show the calculation results for $D_{tr}(T) = 1.0 \times 10^{-20}$ m$^2$/s, which was the value used for the calculation shown in **Figs. 5** and **S10**. The good agreement between experiment and simulation indicates that there is not an appreciable variation in the $D_{tr}(T)$ with film thickness from the vacuum interface to a depth of at least 2 ×15 ML (~10 nm) into the film.

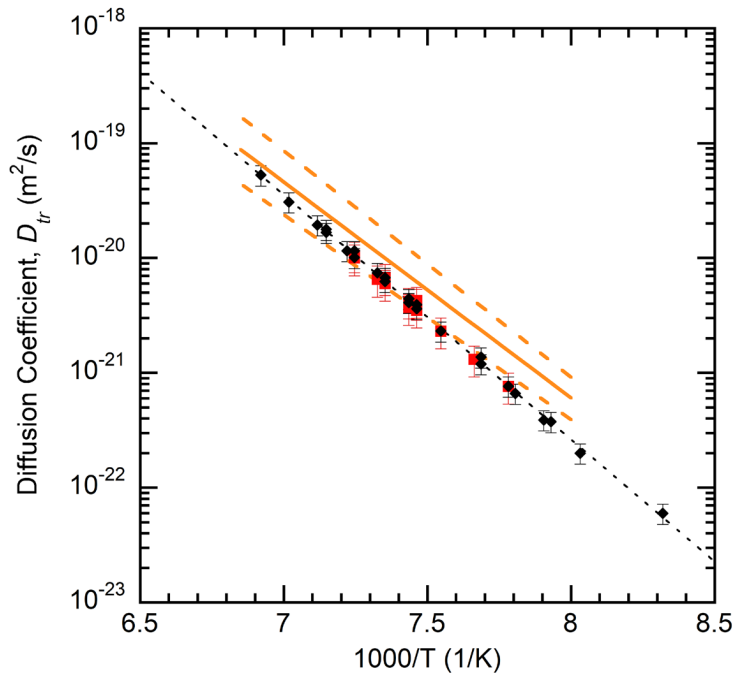

**Fig. S12.** Comparison of $D_{tr}(T)$ vs $1000/T$ reported in the current work (red squares and black diamonds, see **Fig. 6** for details) to the results of recent experiments that investigated diffusion in layered films $H_2^{18}O$ and $H_2^{16}O$ (orange lines).[4] Those experiments measured the desorption rate of $H_2^{18}O$ and $H_2^{16}O$ from films with a total coverage of 100 ML as they were heated with ramp rates from 0.0001 K/s up to 0.01 K/s. The solid orange line is the best fit to those experiments, while the dashed lines show the estimated uncertainty in those results. As discussed in section IV, the diffusive mixing of layered $D_2O/H_2O$ films is likely to slower than the mixing in layered $H_2^{18}O/H_2^{16}O$ due to isotope expects. The results shown here are consistent with these expectations.

**Table 1.** Characteristic times, $\tau(T)$, and diffusion coefficients, $D_{tr}(T)$, for 200 ML water films deposited with a single 20 ML D$_2$O in the middle. The last column shows the diffusion coefficient calculated from the Arrhenius fit to the data shown in Fig. 6: $D_{tr}(T) = D_0 \exp(-E_a/kT)$, where D$_0$ = 2.84 × 10$^{-5}$ m$^2$/s and $E_a$ = 40.76 kJ/mol. The values for $\tau$ listed below were used to scale the integrated IR signals versus time shown as in **Figs. 3b** and **S4b**. **Figure S6** (black diamonds) shows $\tau$ vs $1/T$. The table also lists the values of $D_{tr}$ displayed in Fig. 6 (black diamonds). As discussed in Section S1, converting $\tau(T)$ to $D_{tr}(T)$ was by fixing the value of $D_{tr}(T_g)$ to be 6.25 × 10$^{-21}$ m$^2$/s and $\lambda$ = 3.3 nm for one experiment. That experiment is shown in red below.

| T (K) | $\tau$ (s) | $D_{tr}$ (m$^2$/s) | Arrhenius fit: $D_{tr}$ (m$^2$/s) |
|---|---|---|---|
| 120.2 | 181560 | 6.00E-23 | 5.51E-23 |
| 124.5 | 54470 | 2.00E-22 | 2.25E-22 |
| 126.1 | 28950 | 3.76E-22 | 3.71E-22 |
| 126.5 | 27970 | 3.89E-22 | 4.20E-22 |
| 128.1 | 16390 | 6.64E-22 | 6.81E-22 |
| 128.5 | 14230 | 7.65E-22 | 7.67E-22 |
| 130.1 | 7925 | 1.37E-21 | 1.23E-21 |
| 130.1 | 9080 | 1.20E-21 | 1.23E-21 |
| 132.5 | 4710 | 2.31E-21 | 2.43E-21 |
| 134.0 | 2995 | 3.64E-21 | 3.67E-21 |
| 134.0 | 2770 | 3.93E-21 | 3.67E-21 |
| 134.5 | 2650 | 4.11E-21 | 4.21E-21 |
| 134.5 | 2455 | 4.44E-21 | 4.21E-21 |
| 136.0 | 1610 | 6.76E-21 | 6.29E-21 |
| **136.0** | **1742** | **6.25E-21** | 6.29E-21 |
| 136.5 | 1460 | 7.46E-21 | 7.18E-21 |
| 138.0 | 1080 | 1.01E-20 | 1.06E-20 |
| 138.0 | 940 | 1.16E-20 | 1.06E-20 |
| 138.5 | 942 | 1.16E-20 | 1.21E-20 |
| 139.9 | 653 | 1.67E-20 | 1.72E-20 |
| 139.9 | 613 | 1.78E-20 | 1.72E-20 |
| 140.5 | 560 | 1.94E-20 | 2.00E-20 |
| 142.5 | 353 | 3.08E-20 | 3.26E-20 |
| 144.5 | 206 | 5.29E-20 | 5.24E-20 |

Table 2. Characteristic times, $\tau(T)$, and diffusion coefficients, $D_{tr}(T)$, for experiments measuring the dangling OD signal at the vacuum interface. $\tau(T)$ and $D_{tr}(T)$ are listed for experiments with a 20 ML $D_2O$ film deposited on top of 180 ML $H_2O$ (see **Figs. 4**, **6**, and **S9**). *For experiments where a 20 ML $D_2O$ layer was capped with $H_2O$, only $D_{tr}(T)$ was determined (see **Figs. 5**, **S10**, and **S11** for experiments at 138 K). Comparable experiments with $H_2O$ cap layers were also conducted at 134 and 136 K (not shown).

| $T$ (K) | $\tau$ (s) | $D_{tr}$ (m$^2$/s) | Arrhenius fit: $D_{tr}$ (m$^2$/s) |
|---|---|---|---|
| 128.5 | 14500 | 7.65E-22 | 7.636E-22 |
| 130.5 | 8500 | 1.31E-21 | 1.369E-21 |
| 132.5 | 4800 | 2.31E-21 | 2.413E-21 |
| 134 | 2700 | 4.11E-21 | 3.650E-21 |
| 134 | 2600 | 4.27E-21 | 3.650E-21 |
| 134.5 | 3000 | 3.70E-21 | 4.181E-21 |
| 134.5 | 2640 | 4.20E-21 | 4.181E-21 |
| 136 | 1650 | 6.73E-21 | 6.247E-21 |
| 136 | 1650 | 6.73E-21 | 6.247E-21 |
| 136.5 | 1700 | 6.53E-21 | 7.128E-21 |
| 138 | 1050 | 1.06E-20 | 1.053E-20 |
| 138 | 1050 | 1.06E-20 | 1.053E-20 |
| 134* | | 3.50E-20 | 3.650E-21 |
| 136* | | 6.00E-21 | 6.247E-21 |
| 138* | | 1.00E-21 | 1.053E-20 |

**References**

[1] J. Crank, *The mathematics of diffusion* (Oxford University Press, London, 1975), Second edn., 414.

[2] D. Williams, "Frequency assignments in infra-red spectrum of water," Nature **210**, 194-195 (1966).

[3] A. B. McCoy, "The role of electrical anharmonicity in the association band in the water spectrum," J. Phys. Chem. B **118**, 8286-8294 (2014).

[4] R. S. Smith, W. A. Thornley, G. A. Kimmel, and B. D. Kay, "Supercooled liquid Water diffusivity at temperatures near the glass transition temperature," arXiv:2502.02876 [cond-mat.soft] (2025).